\documentclass{aa}  
\usepackage{graphicx}
\usepackage{txfonts}
\newcommand{\mincir}{\raise -2.truept\hbox{\rlap{\hbox{$\sim$}}\raise5.truept
\hbox{$<$}\ }}
\newcommand{\magcir}{\raise -2.truept\hbox{\rlap{\hbox{$\sim$}}\raise5.truept
\hbox{$>$}\ }}
\newcommand{\siml}{\raise -2.truept\hbox{\rlap{\hbox{$\sim$}}\raise5.truept
\hbox{$<$}\ }}
\newcommand{\simg}{\raise -2.truept\hbox{\rlap{\hbox{$\sim$}}\raise5.truept
\hbox{$>$}\ }}
\newcommand{\be}{\begin{equation}}
\newcommand{\ee}{\end{equation}}
\newcommand{\ba}{\begin{eqnarray}}
\newcommand{\ea}{\end{eqnarray}}
\newcommand {\kpc} {$h_{70}^{-1}$ kpc $\;$}
\newcommand {\kpcc} {$h_{70}^{-1}$ kpc}

\newcommand {\h} {$h_{70}^{-1}$ Mpc$\;$}
\newcommand {\hc} {$h_{100}^{-1}$ Mpc$\;$}
\newcommand {\hh} {$h_{70}^{-1}$ Mpc}
\newcommand {\hhh} {\;h_{70}^{-1} \mathrm{Mpc}}
\newcommand {\ks} {km~s$^{-1} \;$}
\newcommand {\kss} {km~s$^{-1}$}

\newcommand {\lsun} {$h_{70}^{-2}\;L_{\odot} \;$}

\newcommand {\mqui} {$\times 10^{15}\;h_{70}^{-1}\;M_{\odot} \;$}
\newcommand {\mquii} {$\times 10^{15}\;h_{70}^{-1}\;M_{\odot}$}

\newcommand{\degree}{\ensuremath{\mathrm{^\circ}}}
\newcommand{\arcm}{\ensuremath{\mathrm{^\prime}\;}}
\newcommand{\arcs}{\ensuremath{\arcmm\hskip -0.1em\arcmm \;}}
\newcommand{\arcmm}{\ensuremath{\mathrm{^\prime}}}
\newcommand{\arcss}{\ensuremath{\arcmm\hskip -0.1em\arcmm}}
\newcommand{\dotarcs}{\,\rlap{\hbox{$\mathrm{^\prime\hskip-0.1em^\prime}$}}{\hbox{$.$}}\,}

\newcommand{\dotsec}{\,\rlap{\hbox{$\mathrm{^s}$}}{\hbox{$.$}}\,}

\begin{document}
   \title{Internal dynamics of the galaxy cluster Abell 545}  
\subtitle{The ideal case where to study the simultaneous formation of a galaxy
           system and its brightest galaxy}

   \author{
R. Barrena\inst{1,2}
          \and
M. Girardi\inst{3,4}
          \and
W. Boschin\inst{5}
          \and
S. De Grandi\inst{6}
          \and
D. Eckert\inst{7}
          \and
M. Rossetti\inst{7}
}

   \offprints{R. Barrena, \email{rbarrena@iac.es}}

   \institute{ 
     Instituto de
     Astrof\'{\i}sica de Canarias, C/V\'{\i}a L\'actea s/n, E-38205 La
     Laguna (Tenerife), Canary Islands, Spain
\and Departamento de
     Astrof\'{\i}sica, Universidad de La Laguna, Av. del
     Astrof\'{\i}sico Franciso S\'anchez s/n, E-38205 La Laguna
     (Tenerife), Canary Islands, Spain
\and Dipartimento di Fisica dell'Universit\`a degli Studi
     di Trieste - Sezione di Astronomia, via Tiepolo 11, I-34143
     Trieste, Italy
\and INAF -- Osservatorio Astronomico di Trieste,
     via Tiepolo 11, I-34143 Trieste, Italy
\and 
     Fundaci\'on Galileo
     Galilei -- INAF, Rambla Jos\'e Ana Fern\'andez Perez 7, E-38712
     Bre\~na Baja (La Palma), Canary Islands, Spain
\and 
     INAF -- Osservatorio Astronomico di Brera, via E. Bianchi 46,
     I-23807 Merate (LC), Italy
\and 
     INAF -- IASF Milano, via E. Bassini 15, I-20133 Milano, Italy
}

\date{Received  / Accepted }

\abstract{The mechanisms giving rise to diffuse radio emission in
  galaxy clusters, and in particular their connection with cluster
  mergers, are still debated.}{We seek to explore the internal
  dynamics of the radio halo cluster Abell 545. This cluster is also 
  peculiar for hosting in its center a very bright, red, diffuse 
  intracluster light due to an old, stellar population, so bright 
  to be named as ``star pile''.}{Our analysis is mainly based on 
  redshift data for 110 galaxies. We identify 95 cluster members 
  and analyze the cluster internal dynamics by combining galaxy 
  velocities and positions. We also use both photometric and X-ray 
  data.}{We estimate the cluster redshift, $\left<z\right>=0.1580$,
  a large line-of-sight (LOS) velocity dispersion $\sigma_{\rm
    V}\sim 1200$ \kss, and ICM temperature $kT_{\rm X}\sim \,$8
  keV. Our optical and X-ray analyses detect substructures. Optical
  data reveal three main galaxy clumps (one at the
      center hosting the peak of X-ray emission; one at NNW, and
  one at NE); and possibly a fourth clump at South. There is not a
  dominant galaxy and the four brightest galaxies avoid the cluster
  core -- $\gtrsim 0.4$ \h distant from the cluster center -- and 
  are $\gtrsim 1500$ \ks far from the mean cluster velocity. The 
  analysis of the X-ray surface brightness distribution provides 
  us evidence of a disturbed dynamical phase with a likely signature 
  of a shock. Located in the star pile region there is the 
  brightest galaxy of the cluster core (CBCG) and a very compact 
  elliptical galaxy. We show that the star pile, which
  has a previously determined redshift, has a similar redshift
  to that of the CBCG. Both the star pile and the CBCG are at rest in
  the cluster rest frame.  The elongation of the star pile and its
  relative position with respect to the CBCG indicate the same
  direction pointed out by the NE clump.}{The emerging picture
  of Abell 545 is that of a massive, $M(R<1.6 \hhh)=(1.1$--$1.8)$
  \mquii, very complex cluster with merging occurring along two
  directions.  Abell 545 gives another proof in the favor of the
  connection between cluster merger and extended, diffuse radio
  emission.  The star pile, likely due to the process of a brightest
  galaxy forming in the cluster core. Abell 545 represents a textbook cluster
  where to study the simultaneous formation of a galaxy system and its
  brightest galaxy.}


   \maketitle
%

\section{Introduction}
\label{intr}

Merging processes constitute an essential ingredient of the evolution
of galaxy clusters (see Feretti et al. \cite{fer02b} for a review). An
interesting aspect of these phenomena is the possible connection
between cluster mergers and extended, diffuse radio sources: halos and
relics. The synchrotron radio emission of these sources demonstrates
the existence of large-scale cluster magnetic fields and of
widespread relativistic particles. Cluster mergers have been proposed
to provide the large amount of energy necessary for electron
re-acceleration to relativistic energies and for magnetic field
amplification (Tribble \cite{tri93}; Feretti \cite{fer99}; Feretti
\cite{fer02a}; Sarazin \cite{sar02}). Radio relics (``radio gischts''
as referred to by Kempner et al. \cite{kem04}), which are polarized
and elongated radio sources located in the cluster peripheral regions,
seem to be directly associated with merger shocks (e.g., Ensslin et
al. \cite{ens98}; Roettiger et al. \cite{roe99}; Ensslin \&
Gopal-Krishna \cite{ens01}; Hoeft et al. \cite{hoe04}).  Radio halos
are sources that permeate the cluster volume in a similar way to
the X-ray emitting gas of the intra-cluster medium (hereafter
ICM) and are usually found unpolarized (but see Govoni et
  al. \cite{gov05}). Radio halos are more likely to be associated
with the turbulence following a cluster merger although the precise
radio formation scenario remains unclear (re-acceleration vs. hadronic
models e.g., Brunetti et al.  \cite{bru09}; Ensslin et al.
\cite{ens11}).  Very recently, a unified halo-relic model has
  been presented in the framework of hadronic models where the
  time-dependence of the magnetic fields and of the cosmic ray
  distributions is taken into account to explain both halos and (most)
  relics (Keshet \cite{kes10}). In this model the ICM magnetization is
  triggered by a merger event, in part but probably not exclusively in
  the wake of merger shocks.

The study of galaxy clusters with radio emissions offers a unique
  tool to estimate strength and structure of large-scale magnetic
  fields and  might have important cosmological implications
  (see Dolag et al. \cite{dol04} and Ferrari et al. \cite{fer08} for
  recent reviews). If it will be confirmed and understood the
  connection between radio halos/relics and cluster mergers, one could
  use radio diffuse sources to directly follow the formation of the
large scale structure.  The study of clusters with radio halos/relics will
  likely contribute to quantify the effect of the non-thermal pressure
  to the estimate of mass and temperature in galaxy clusters (e.g.,
  Loeb \& Mao \cite{loe94}; Dolag \& Schindler \cite{dol00};
  Markevitch \cite{mar10}) and, more in general, the thermal and non
  thermal effects of cluster mergers on global properties and
  cosmological parameters (e.g., Sarazin \cite{sar04}).

Unfortunately, one has been able to study these phenomena only
recently on the basis of a sufficient statistics, i.e.  few dozen
clusters hosting diffuse radio sources up to $z\sim 0.4$
(e.g., Giovannini et al. \cite{gio99}; see also Giovannini \& Feretti
\cite{gio02}; Feretti \cite{fer05}; Giovannini et al. \cite{gio09})
with MACS J0717.5+3745 at $z=0.55$ being the most distant one (Bonafede
et al. \cite{bon09}). This allowed a few authors to attempt a
classification (e.g., Kempner et al. \cite{kem04}; Ferrari et
al. \cite{fer08}). It is expected that new radio telescopes
will largely increase the statistics of diffuse sources (e.g., LOFAR,
Cassano et al. \cite{cas10a}) and allow the study of 
  diffuse radio emission in low X-ray luminosity clusters to
discriminate among theories of halo formation (e.g., Cassano et
al. \cite{cas05}; Ensslin et al. \cite{ens11}).

From the observational point of view, there is growing evidence of the
connection between diffuse radio emission and cluster mergers, since
up to now diffuse radio sources have been detected only in merging
systems (see, e.g., Cassano et al. \cite{cas10b}). In most cases the
cluster dynamical state has been derived from X-ray observations (see
Buote \cite{buo02}; Cassano et al.  \cite{cas10b}).  X-ray studies of
merging clusters are useful to discriminate between pre- and
post-collision phases and to give useful information about the
behavior of the ICM, in particular in the case of the detection of a
shock. In fact, one can compute the shock Mach number from X-ray data,
infer the shock wave velocity and the age of the merger, compare the
dissipated energy with that required by the radio source (e.g.,
Sarazin \cite{sar02}; Markevitch \cite{mar02}; Finoguenov et
al. \cite{fin10}). This contributes to constrain the magnetic field
and/or discriminate between models (e.g., Finoguenov et
al. \cite{fin10}).

Optical data are a powerful way to investigate the presence and the
dynamics of cluster mergers, too (e.g., Girardi \& Biviano
\cite{gir02}). The spatial and kinematical analysis of member galaxies
allow us to reveal and measure the amount of substructure, and to
detect and analyze possible pre-merging clumps or merger remnants.
This optical information is really complementary to X-ray information
since galaxies and the intracluster medium react on different
timescales during a merger (see, e.g., numerical simulations by
Roettiger et al. \cite{roe97}).  In this context, we are conducting an
intensive observational and data analysis program to study the
internal dynamics of clusters with diffuse radio emission by using
member galaxies (DARC - Dynamical Analysis of Radio Clusters -
  project, see Girardi et al. \cite{gir07}\footnote{see also the web
  site of the DARC project
  http://adlibitum.oat.ts.astro.it/girardi/darc.}). The analysis
  based on the position/kinematics of member galaxies has the
  advantage that galaxies and dark matter (DM, i.e. the 70-80\% of
  cluster mass) have similar spatial distributions in both quiet and
  substructured clusters as show by weak gravitational lensing data
  (e.g., Kneib et al. \cite{kne93}; Okabe \& Umetsu \cite{oka08}), in
  agreement with the fact that both galaxies and DM can be considered
  non collisional cluster components.  The interesting consequence is
  that one can directly estimate (projected) positions velocities, and
  masses of subclusters using galaxies as tracer particles.  Combined
  X-ray and optical information is exploited at their best in ad hoc
  numerical simulations of merging clusters where dark matter, gas and
  shock behaviors can be analyzed in function of the merger age (e.g.,
  Springel \& Farrar \cite{spr07}; Mastropietro \& Burkert
  \cite{mas08}).  However, for clusters formed by two main
  subclusters, one can use the optical observational information in
  the approximation of the analytical two-body model (Beers et
  al. \cite{bee82}). The angle of view of the merger and the merger
  age are the two unknown quantities for which additional information
  can be used. This additional information is, e.g., the angle of view
  suggested by the relative efficiency of the substructure methods
  (Pinkney et al. \cite{pin96}); the time elapsed since the subcluster
  collision as suggested by the X-ray temperature variation during the
  merger (Ricker \& Sarazin \cite{ric01}) or as inferred from the shock
  Mach number derived from X-ray or radio data (see, e.g., Boschin et
  al. \cite{gir10}; Girardi et al. \cite{gir10} for application of the
  bimodal model in DARC clusters).

As for the DARC project, we found evidence of major optical
  substructure for all the twelve clusters already analyzed on the
  base of enough statistics (60--170 members).  These twelve
  clusters span a wide range of optical morphologies (showing two,
  three, many important clumps), view angles, and radio morphologies
  (double relics, relics, halos).  This is the main reason why we are
  studying several individual clusters in order to face on a global
  classification and present a meaningful, statistical study. Most
  clusters exhibiting diffuse radio emission have a relatively high
  gravitational mass (higher than $0.7\times 10^{15}$ within $2$ \hh;
  see Giovannini \& Feretti \cite{gio02}) and, indeed, most DARC
  clusters are very massive clusters with few exceptions (Boschin et
  al. \cite{bos08}).  

A comprehensive statistical analysis of the whole DARC sample and
  the conclusions derived on the origin of halos/relics will be the
  subject of a future study. Here we focalize our study on the
  internal dynamics of Abell 545. In particular, we show as this
  cluster can be used to check models of the origin of halos/relics,
  although we stress that only the analysis on a large sample is
  expected to be really advantageous.

\subsection{Abell 545}

During our observational program, we have conducted an intensive study
of the cluster \object{Abell 545} (hereafter A545).  A545 is a very
rich, X-ray luminous cluster: Abell richness class $=4$ (the second
richest cluster in the Abell catalog, Abell et al. \cite{abe89});
$L_\mathrm{X}$(0.1-2.4 keV)=5.67$\times 10^{44}$ $h_{70}^{-2}$ Mpc
erg\ s$^{-1}$ (B\"ohringer et al. \cite{boe04}) and $kT_{\rm
  X}=5.5^{+5.5}_{-2.1}$ keV (David et al. \cite{dav93}).

Optically, the cluster is
classified as Bautz-Morgan class III (Abell et al. \cite{abe89}) and
has a Rood-Sastry morphological type ``I$_{\rm c}$'', i.e. clumpy irregular
(Struble \& Ftaclas \cite{str94}). According to the spectra for two
luminous galaxies A545 lies at $z=0.152$ (Schneider et
al. \cite{sch83}).

A545 is peculiar for hosting a very bright diffuse light, so bright to
be named as ``star pile'' by Struble (\cite{str88}) who
  discovered a red, elongated low surface density feature located
  close to the cluster center of A545 on both the E and O plates of
  the Palomar Observatory Sky Survey. Struble interpreted it as
  intracluster matter.  Based on archival VLT-images and long-slit
spectra obtained with Gemini-GMOS, Salinas et al. (\cite{sal07},
hereafter S07) found that the star pile is indeed associated with
A545, its velocity being $\sim$ 1300 \ks higher than that of a faint
central galaxy. The spectra indicate an old, presumably metal-rich
population. Its brightness profile is much shallower than that of
typical cD-galaxies. However, as recognized by S07, the formation of
this star pile remains elusive, until a dynamical analysis of the
cluster itself become available.

Evidence for an unrelaxed state of A545 comes from X-ray data
analyses.  The X-ray structure of this cluster is elongated as
reported by Buote \& Tsai (\cite{buo96}, from ROSAT data) who selected
A545 as ``reference'' cluster for elliptical morphology. In
particular, they found that A545 is the only cluster in their sample
of 59 clusters that is highly elongated but does not display any
obvious center offset, i.e., a center displacement from the outer
  emission (Jones \& Forman \cite{jon92}).  An image from the XMM
archive also shows that A545 is strongly NNW-SSE elongated (see Fig.~6
of S07 and our analysis in Sect.~4). Its unrelaxed state is consistent
with the absence of a cooling flow (White et al. \cite{whi97} from
{\it Einstein} data). To date, the possibly involved subclusters
  have not been yet detected and there is no information about the merger
  age.  Indeed, the internal dynamical of A545 has never been
  analyzed.

The presence of a giant radio halo in A545 was suggested by Giovannini
et al. (\cite{gio99}) and then the halo was studied by Bacchi et
al. (\cite{bac03}). The halo structure is centrally located and rather
regular (slightly elongated in the NE-SW direction).  The radio power
of A545 accommodates well on the relation between radio power and
dipole power ratio (Buote \cite{buo01}) and on the relation between
radio power and X-ray luminosity (Bacchi et al. \cite{bac03}).

We included this cluster in our DARC sample and obtained new
spectroscopic and photometric data from the Telescopio Nazionale
Galileo (TNG) and the Isaac Newton Telescope (INT), respectively. Our
present analysis is based on these optical data and XMM-Newton Science
Archive data, too.  

This paper is organized as follows. We present our new optical data
and the cluster catalog in Sect.~2. We present our results about the
cluster structure based on optical and X-ray data in Sects.~3 and 4,
respectively.  We discuss our results and present our conclusions
in Sect.~5.

\begin{figure*}
\centering 
\includegraphics[width=18cm]{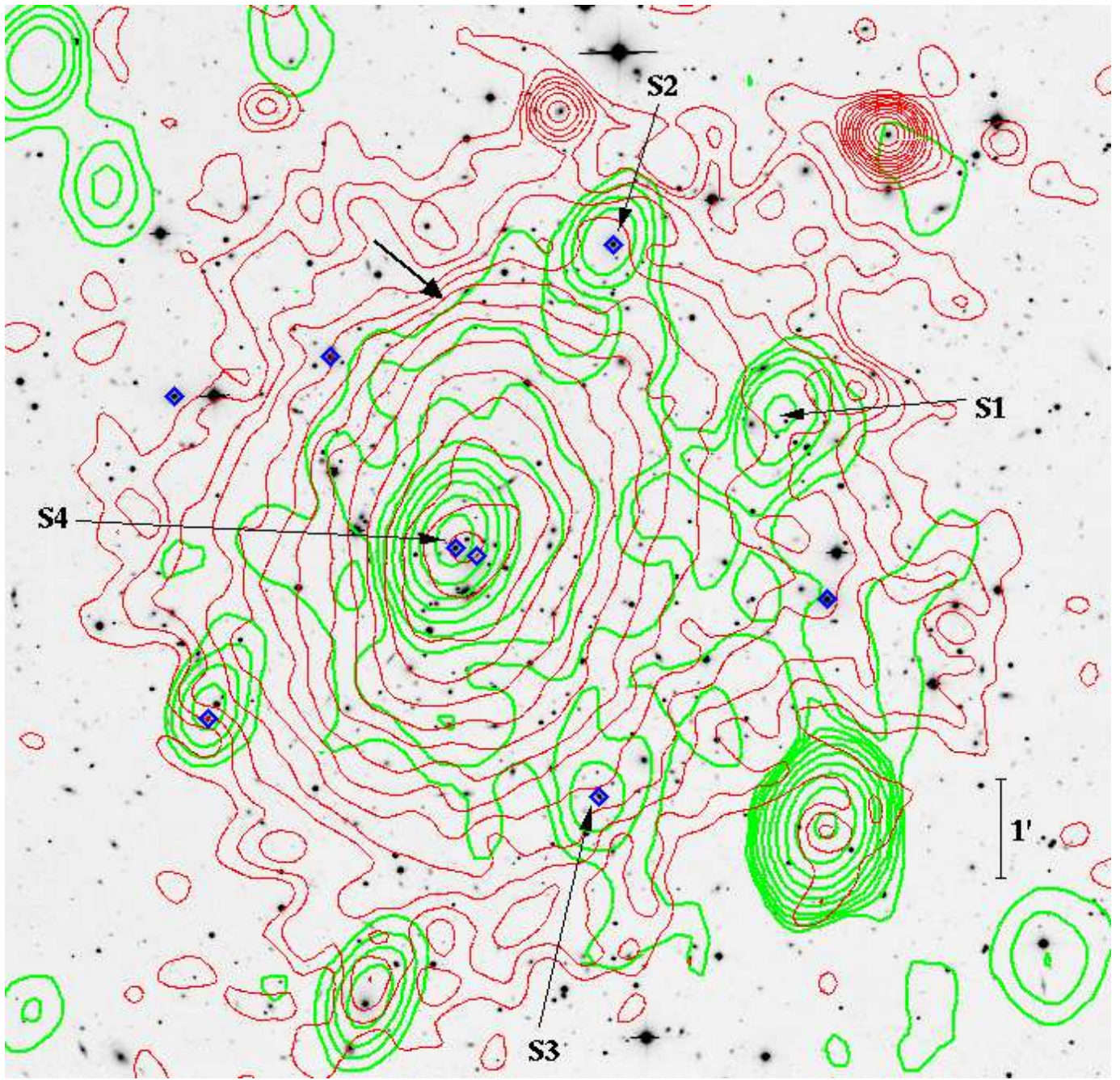}
\caption{INT $r^{\prime}$-band image of the cluster A545 (north at
  the top and east to the left) with, superimposed, the contour levels
  of the XMM archival image (ID.~0304750101) in the 0.4-2 keV energy
  range (thin/red contours) and the contour levels of a VLA radio
  image at 1.4 GHz (thick/green contours, see Bacchi et
  al. \cite{bac03}). Blue rotate squares highlight the positions of
  bright galaxies cited in the text. Labels S1, S2, S3 and S4 indicate
  the positions of radio sources listed by Bacchi et al..  In
    particular, the galaxy indicated by ``S4'' is the brightest galaxy in
    the cluster central region (CBCG).  In the northern region, an
  arrow highlights a sharp discontinuity in the X-ray surface
  brightness (see text).}
\label{figimage}
\end{figure*}
 
Unless otherwise stated, we indicate errors at the 68\% confidence
level (hereafter c.l.).  Throughout this paper, we use $H_0$=70 
km s$^{-1}$ Mpc$^{-1}$ and $h_{70}=H_0/(70$ km s$^{-1}$ Mpc$^{-1}$)
in a flat cosmology with $\Omega_0=0.3$ and $\Omega_{\Lambda}=0.7$. 
In the adopted cosmology, 1\arcm corresponds to $\sim 164$ \kpc at 
the cluster redshift. 

\begin{figure*}
\centering 
\includegraphics[width=18cm]{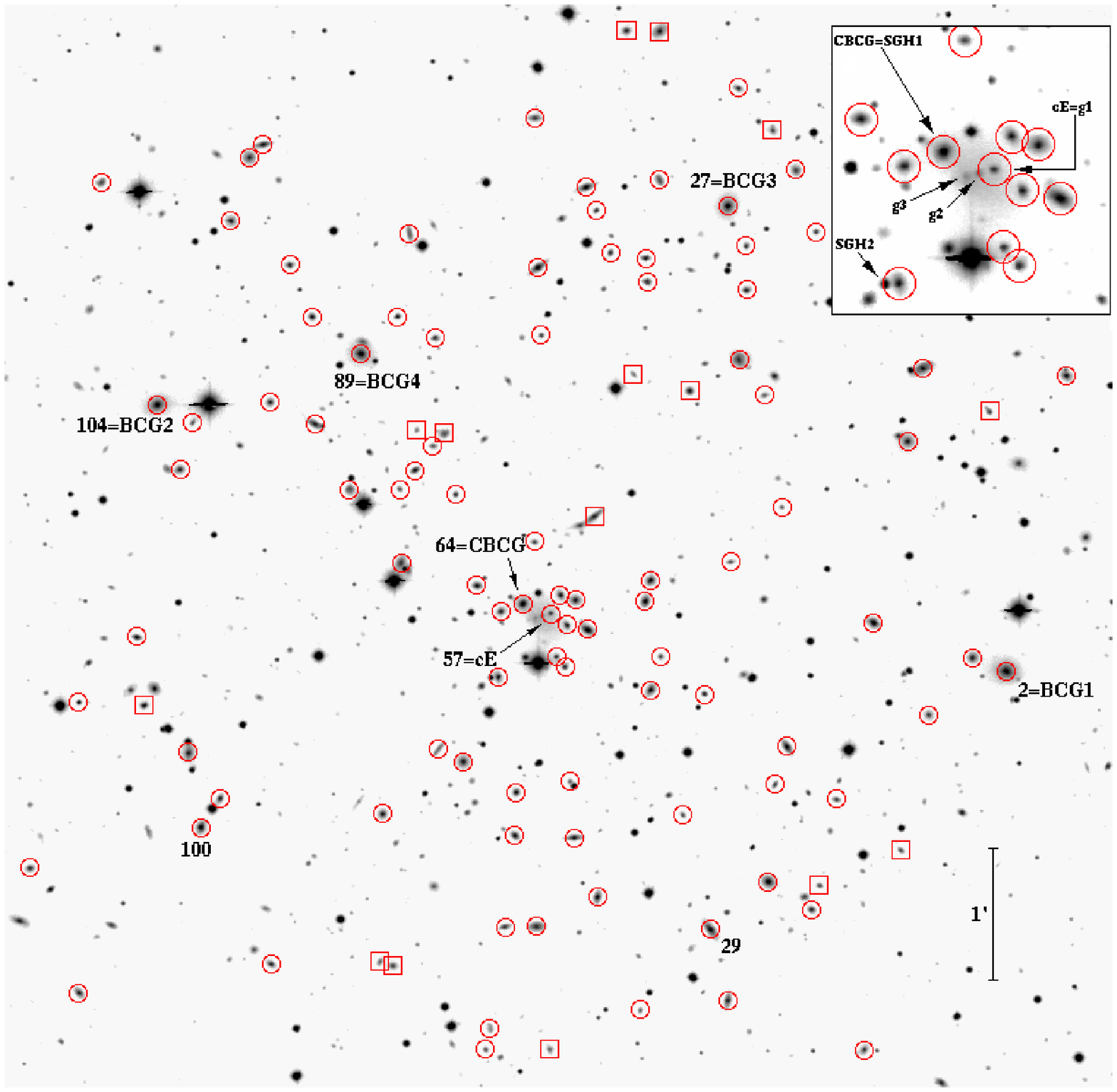}
\caption{INT $r^{\prime}$-band image of the cluster A545 (north at
  the top and east to the left). Circles and squares indicate 
    spectroscopically cluster members and non-members, respectively
  (see Table~\ref{catalogA545}). Labels indicate the IDs of cluster
  galaxies cited in the text. In the insect figure the corresponding 
IDs of S07 are given, too.}
\label{figottico}
\end{figure*}

\section{New data and galaxy catalog}
\label{data}

Multi-object spectroscopic observations of A545 were carried out at
the TNG telescope in October 2009. We used DOLORES/MOS with the LR-B
Grism 1, yielding a dispersion of 187 \AA/mm.  We used the new
$2048\times2048$ pixels E2V CCD, with a pixel size of 13.5 $\mu$m. In
total, we observed 4 MOS masks for a total of 142 slits. We acquired
three exposures of 1200 s for each mask.  Wavelength calibration was
performed using helium and neon-mercury lamps. Reduction of
spectroscopic data was carried out using the IRAF\footnote{IRAF is
  distributed by the National Optical Astronomy Observatories, which
  are operated by the Association of Universities for Research in
  Astronomy, Inc., under cooperative agreement with the National
  Science Foundation.} package. Radial velocities were determined
using the cross-correlation technique (Tonry \& Davis \cite{ton79})
implemented in the RVSAO package (developed at the Smithsonian
Astrophysical Observatory Telescope Data Center). Each spectrum was
correlated against five templates for a variety of galaxy spectral
types: E, S0, Sa, Sb, and Sc (Kennicutt \cite{ken92}). The template
producing the highest value of $\cal R$, i.e., the parameter given by
RVSAO and related to the signal-to-noise ratio of the correlation
peak, was chosen. Moreover, all spectra and their best correlation
functions were examined visually to verify the redshift
determination. In seven cases (IDs.~3, 31, 34, 42, 71, 72, and 77; see
Table~\ref{catalogA545}), we assumed the EMSAO redshift to be a
reliable estimate of the redshift.

Our spectroscopic catalog lists 110 galaxies in the field of A545.


\begin{table}[!ht]
        \caption[]{Velocity catalog of 110 spectroscopically measured
          galaxies in the field of the cluster A545.}
         \label{catalogA545}
              $$ 
           \begin{array}{r c c c c r r c}
            \hline
            \noalign{\smallskip}
            \hline
            \noalign{\smallskip}

\mathrm{ID} & \mathrm{\alpha}(\mathrm{J}2000),\mathrm{\delta}(\mathrm{J}2000)& g^{\prime}& r^{\prime}& i^{\prime}&\mathrm{v}\,\,\,\,\,\,\, & \mathrm{\Delta}\mathrm{v}& \mathrm{Mem.}\\
& \mathrm{h:m:s,\degree:\arcmm:\arcs}& & & &\mathrm{(\,km}&\mathrm{s^{-1}\,)}& \\

            \hline
            \noalign{\smallskip}  
 1         & 05\ 32\ 09.06 , -11\ 30\ 52.4& 18.72&17.82&  17.53& 46877& 33& 1 \\
 2         & 05\ 32\ 10.89 , -11\ 33\ 05.9& 17.05&16.13&  15.89& 48992& 35& 1 \\
 3         & 05\ 32\ 11.41 , -11\ 31\ 08.3& 19.92&19.22&  19.01&103532&100& 0 \\
 4         & 05\ 32\ 11.93 , -11\ 32\ 59.8& 18.64&17.95&  17.60& 47906& 35& 1 \\
 5         & 05\ 32\ 13.27 , -11\ 33\ 25.4& 19.61&18.70&  18.37& 45015& 39& 1 \\
 6         & 05\ 32\ 13.47 , -11\ 30\ 48.6& 18.05&17.33&  17.04& 45364& 40& 1 \\
 7         & 05\ 32\ 13.93 , -11\ 31\ 21.8& 19.30&18.59&  18.33& 46967& 38& 1 \\
 8         & 05\ 32\ 14.13 , -11\ 34\ 26.5& 20.61&19.35&  19.03& 72394& 55& 0 \\
 9         & 05\ 32\ 14.98 , -11\ 32\ 43.6& 18.14&17.50&  17.24& 44558& 34& 1 \\
10         & 05\ 32\ 15.26 , -11\ 35\ 56.8& 19.56&18.58&  18.23& 47521& 56& 1 \\
11         & 05\ 32\ 16.11 , -11\ 34\ 03.4& 20.22&19.29&  18.92& 47956& 72& 1 \\
12         & 05\ 32\ 16.64 , -11\ 34\ 42.3& 20.74&19.46&  19.01& 81561& 45& 0 \\
13         & 05\ 32\ 16.75 , -11\ 29\ 46.7& 20.31&19.44&  19.06& 47925& 64& 1 \\
14         & 05\ 32\ 16.88 , -11\ 34\ 53.3& 19.40&18.71&  18.40& 46953& 42& 1 \\
15         & 05\ 32\ 17.37 , -11\ 29\ 18.7& 19.29&18.43&  18.13& 47329& 56& 1 \\
16         & 05\ 32\ 17.64 , -11\ 33\ 39.4& 18.70&17.77&  17.41& 48957& 35& 1 \\
17         & 05\ 32\ 17.79 , -11\ 31\ 51.1& 20.99&20.14&  19.81& 48992& 59& 1 \\
18         & 05\ 32\ 18.01 , -11\ 33\ 56.6& 20.28&19.46&  19.12& 47164& 64& 1 \\
19         & 05\ 32\ 18.08 , -11\ 29\ 00.7& 19.91&18.99&  18.62& 76942& 62& 0 \\
20         & 05\ 32\ 18.21 , -11\ 34\ 40.5& 18.21&17.26&  16.88& 47765& 42& 1 \\
21         & 05\ 32\ 18.33 , -11\ 31\ 00.6& 20.26&19.45&  19.12& 48803& 60& 1 \\
22         & 05\ 32\ 18.88 , -11\ 30\ 12.8& 19.52&18.63&  18.27& 46794& 43& 1 \\
23         & 05\ 32\ 18.89 , -11\ 29\ 52.8& 20.05&19.22&  18.90& 47230& 64& 1 \\
24         & 05\ 32\ 19.11 , -11\ 30\ 44.2& 17.92&17.31&  17.02& 48594& 72& 1 \\
25         & 05\ 32\ 19.14 , -11\ 28\ 41.5& 19.76&18.85&  18.45& 46621& 40& 1 \\
26         & 05\ 32\ 19.37 , -11\ 32\ 15.7& 20.66&19.68&  19.35& 49423&107& 1 \\
27         & 05\ 32\ 19.45 , -11\ 29\ 34.8& 17.70&16.73&  16.36& 49449& 38& 1 \\
28         & 05\ 32\ 19.45 , -11\ 35\ 34.0& 19.01&18.21&  17.79& 46433& 48& 1 \\
29         & 05\ 32\ 20.00 , -11\ 35\ 02.0& 18.12&17.41&  17.11& 45708& 40& 1 \\
30         & 05\ 32\ 20.18 , -11\ 33\ 15.5& 20.16&19.20&  18.80& 46677& 59& 1 \\
31         & 05\ 32\ 20.63 , -11\ 30\ 58.2& 19.12&18.50&  18.20& 68252&116& 0 \\
32         & 05\ 32\ 20.83 , -11\ 34\ 10.0& 20.53&19.65&  19.25& 47072& 73& 1 \\
33         & 05\ 32\ 21.52 , -11\ 32\ 58.5& 20.34&19.45&  19.09& 45745& 88& 1 \\
34         & 05\ 32\ 21.56 , -11\ 28\ 15.6& 18.54&17.73&  17.35& 67715&100& 0 \\
35         & 05\ 32\ 21.56 , -11\ 29\ 23.0& 19.30&18.64&  18.37& 50360& 44& 1 \\
36         & 05\ 32\ 21.83 , -11\ 33\ 13.4& 18.90&17.98&  17.60& 48049& 25& 1 \\
37         & 05\ 32\ 21.84 , -11\ 32\ 24.0& 18.91&17.96&  17.57& 47452& 29& 1 \\
38         & 05\ 32\ 21.94 , -11\ 30\ 08.9& 19.71&18.76&  18.34& 45527& 73& 1 \\
39         & 05\ 32\ 21.98 , -11\ 29\ 58.1& 19.51&18.68&  18.32& 46272& 49& 1 \\
40         & 05\ 32\ 21.99 , -11\ 32\ 33.2& 18.56&17.82&  17.52& 45652& 22& 1 \\
41         & 05\ 32\ 22.14 , -11\ 35\ 38.2& 20.52&19.99&  19.91& 50629& 61& 1 \\
42         & 05\ 32\ 22.36 , -11\ 30\ 50.6& 20.46&19.93&  19.59&116937&100& 0 \\
43         & 05\ 32\ 22.58 , -11\ 28\ 15.2& 19.64&18.45&  17.98& 67832& 73& 0 \\
44         & 05\ 32\ 23.06 , -11\ 29\ 55.7& 20.03&19.05&  18.68& 49512& 52& 1 \\
45         & 05\ 32\ 23.46 , -11\ 34\ 46.8& 19.10&18.18&  17.82& 47744& 44& 1 \\
46         & 05\ 32\ 23.49 , -11\ 29\ 36.6& 20.29&19.39&  19.02& 46560& 52& 1 \\
47         & 05\ 32\ 23.55 , -11\ 31\ 54.8& 18.73&17.97&  17.53& 41964& 51& 0 \\
48         & 05\ 32\ 23.77 , -11\ 32\ 46.0& 18.22&17.27&  16.89& 46105& 22& 1 \\
49         & 05\ 32\ 23.82 , -11\ 29\ 25.8& 18.61&17.73&  17.31& 44975& 38& 1 \\
50         & 05\ 32\ 24.14 , -11\ 32\ 32.6& 19.02&18.05&  17.63& 48736& 42& 1 \\
51         & 05\ 32\ 24.18 , -11\ 34\ 20.0& 19.01&18.12&  17.74& 47044& 42& 1 \\
52         & 05\ 32\ 24.30 , -11\ 33\ 54.8& 21.77&20.64&  20.33& 46653&109& 1 \\
53         & 05\ 32\ 24.41 , -11\ 32\ 44.0& 19.51&18.75&  18.51& 48216& 66& 1 \\
54         & 05\ 32\ 24.46 , -11\ 33\ 02.8& 19.57&18.55&  18.46& 48636& 40& 1 \\
55         & 05\ 32\ 24.60 , -11\ 32\ 30.4& 19.26&18.32&  17.86& 47212& 35& 1 \\
               
                        \noalign{\smallskip}			    
            \hline					    
            \noalign{\smallskip}			    
            \hline					    
         \end{array}
     $$ 
         \end{table}
\addtocounter{table}{-1}
\begin{table}[!ht]
          \caption[ ]{Continued.}
     $$ 
           \begin{array}{r c c c c r r c}
            \hline
            \noalign{\smallskip}
            \hline
            \noalign{\smallskip}

\mathrm{ID} & \mathrm{\alpha}(\mathrm{J}2000),\mathrm{\delta}(\mathrm{J}2000)& g^{\prime}& r^{\prime}& i^{\prime}&\mathrm{v}\,\,\,\,\,\,\, & \mathrm{\Delta}\mathrm{v}& \mathrm{Mem.}\\
& \mathrm{h:m:s,\degree:\arcmm:\arcs}& & & &\mathrm{(\,km}&\mathrm{s^{-1}\,)}& \\

            \hline
            \noalign{\smallskip}
56          & 05\ 32\ 24.73  , -11\ 32\ 58.3& 19.58& 18.53& 18.27& 48907&  74& 1 \\
57          & 05\ 32\ 24.90  , -11\ 32\ 38.7& 20.88& 19.97& 19.54& 46354& 109& 1 \\
58          & 05\ 32\ 24.93  , -11\ 35\ 55.6& 21.02& 19.34& 18.80&117147&  90& 0 \\
59          & 05\ 32\ 25.20  , -11\ 30\ 32.7& 20.30& 19.45& 19.11& 46321& 153& 1 \\
60          & 05\ 32\ 25.28  , -11\ 30\ 02.0& 18.54& 17.60& 17.18& 47505&  35& 1 \\
61          & 05\ 32\ 25.33  , -11\ 35\ 00.0& 18.74& 17.91& 17.59& 48896&  51& 1 \\
62          & 05\ 32\ 25.39  , -11\ 28\ 54.4& 19.31& 18.52& 18.09& 49441&  53& 1 \\
63          & 05\ 32\ 25.40  , -11\ 32\ 06.2& 20.20& 19.35& 18.96& 45711&  87& 1 \\
64          & 05\ 32\ 25.76  , -11\ 32\ 34.2& 18.31& 17.29& 16.71& 47071&  26& 1 \\
65          & 05\ 32\ 25.96  , -11\ 33\ 59.4& 19.46& 18.55& 18.17& 46995&  36& 1 \\
66          & 05\ 32\ 26.00  , -11\ 34\ 19.0& 19.36& 18.44& 18.02& 47219&  38& 1 \\
67          & 05\ 32\ 26.30  , -11\ 35\ 00.3& 19.69& 18.81& 18.48& 45362&  87& 1 \\
68          & 05\ 32\ 26.30  , -11\ 35\ 00.3& 19.69& 18.81& 18.48& 49434&  44& 1 \\
69          & 05\ 32\ 26.43  , -11\ 32\ 37.7& 18.89& 17.92& 17.41& 46297&  35& 1 \\
70          & 05\ 32\ 26.52  , -11\ 33\ 07.1& 19.77& 18.97& 18.72& 48242&  44& 1 \\
71          & 05\ 32\ 26.76  , -11\ 35\ 46.0& 19.86& 19.56& 19.56& 47480& 149& 1 \\
72          & 05\ 32\ 26.90  , -11\ 35\ 55.6& 20.20& 19.77& 19.52& 47359& 320& 1 \\
73          & 05\ 32\ 27.17  , -11\ 32\ 25.8& 19.29& 18.35& 17.93& 47486&  39& 1 \\
74          & 05\ 32\ 27.58  , -11\ 33\ 45.5& 18.65& 17.71& 17.33& 48020&  43& 1 \\
75          & 05\ 32\ 27.82  , -11\ 31\ 44.7& 20.15& 19.37& 19.05& 46074&  86& 1 \\
76          & 05\ 32\ 28.18  , -11\ 31\ 16.9& 19.39& 18.65& 18.42& 97676&  49& 0 \\
77          & 05\ 32\ 28.35  , -11\ 33\ 39.5& 19.59& 19.03& 18.67& 45136& 100& 1 \\
78          & 05\ 32\ 28.46  , -11\ 30\ 33.6& 19.94& 19.01& 18.63& 45406&  51& 1 \\
79          & 05\ 32\ 28.52  , -11\ 31\ 22.5& 20.37& 19.53& 19.16& 50551&  52& 1 \\
80          & 05\ 32\ 29.03  , -11\ 31\ 15.4& 21.04& 20.27& 19.91& 62889& 240& 0 \\
81          & 05\ 32\ 29.06  , -11\ 31\ 33.5& 19.42& 18.52& 18.12& 45954&  38& 1 \\
82          & 05\ 32\ 29.25  , -11\ 29\ 46.4& 19.25& 18.64& 18.37& 46892& 101& 1 \\
83          & 05\ 32\ 29.50  , -11\ 32\ 15.6& 19.23& 18.88& 18.45& 45963&  56& 1 \\
84          & 05\ 32\ 29.55  , -11\ 31\ 42.1& 20.46& 19.70& 19.47& 47421&  52& 1 \\
85          & 05\ 32\ 29.60  , -11\ 30\ 24.1& 19.94& 18.98& 18.61& 49406&  64& 1 \\
86          & 05\ 32\ 29.75  , -11\ 35\ 17.4& 20.18& 18.75& 18.20& 97303&  35& 0 \\
87          & 05\ 32\ 30.06  , -11\ 34\ 08.8& 19.22& 18.34& 17.97& 46322&  60& 1 \\
88          & 05\ 32\ 30.14  , -11\ 35\ 15.6& 20.71& 19.41& 18.81& 97292&  70& 0 \\
89          & 05\ 32\ 30.74  , -11\ 30\ 40.7& 17.78& 16.81& 16.39& 49403&  31& 1 \\
90          & 05\ 32\ 31.09  , -11\ 31\ 42.1& 18.83& 18.40& 18.17& 46229&  64& 1 \\
91          & 05\ 32\ 32.16  , -11\ 31\ 12.4& 18.59& 17.99& 17.62& 49155&  51& 1 \\
92          & 05\ 32\ 32.23  , -11\ 30\ 23.9& 19.49& 18.55& 18.21& 48119&  31& 1 \\
93          & 05\ 32\ 32.92  , -11\ 30\ 00.3& 19.77& 18.80& 18.41& 49526&  42& 1 \\
94          & 05\ 32\ 33.50  , -11\ 35\ 16.4& 19.59& 18.71& 18.21& 47562&  51& 1 \\
95          & 05\ 32\ 33.52  , -11\ 31\ 02.4& 19.67& 18.73& 18.37& 48801&  60& 1 \\
96          & 05\ 32\ 33.74  , -11\ 29\ 05.8& 18.87& 18.04& 17.72& 48184&  52& 1 \\
97          & 05\ 32\ 34.16  , -11\ 29\ 11.8& 18.81& 17.94& 17.59& 48742&  38& 1 \\
98          & 05\ 32\ 34.73  , -11\ 29\ 40.2& 19.24& 18.33& 17.95& 46204&  44& 1 \\
99          & 05\ 32\ 35.07  , -11\ 34\ 01.5& 19.50& 18.59& 18.19& 47544&  47& 1 \\
100         & 05\ 32\ 35.65  , -11\ 34\ 14.5& 18.45& 17.54& 17.18& 47375&  34& 1 \\
101         & 05\ 32\ 35.90  , -11\ 31\ 11.3& 20.40& 19.48& 19.11& 47906&  77& 1 \\
102         & 05\ 32\ 36.04  , -11\ 33\ 40.6& 18.62& 17.67& 17.27& 46751&  52& 1 \\
103         & 05\ 32\ 36.29  , -11\ 31\ 32.5& 19.30& 18.50& 18.18& 46196&  65& 1 \\
104         & 05\ 32\ 37.00  , -11\ 31\ 03.3& 17.58& 16.62& 16.30& 48721&  30& 1 \\
105         & 05\ 32\ 37.40  , -11\ 33\ 19.1& 19.45& 18.48& 18.10& 55643&  47& 0 \\
106         & 05\ 32\ 37.63  , -11\ 32\ 48.2& 19.39& 18.56& 18.23& 46266&  34& 1 \\
107         & 05\ 32\ 38.73  , -11\ 29\ 22.9& 19.46& 18.85& 18.52& 47549&  72& 1 \\
108         & 05\ 32\ 39.41  , -11\ 33\ 17.5& 20.07& 18.98& 18.41& 45794&  96& 1 \\
109         & 05\ 32\ 39.41  , -11\ 35\ 28.9& 19.40& 18.46& 18.05& 45749&  57& 1 \\
110         & 05\ 32\ 40.91  , -11\ 34\ 32.5& 19.69& 18.73& 18.31& 48224& 122& 1 \\
                        \noalign{\smallskip}                        
            \hline                                          
            \noalign{\smallskip}                            
            \hline                                          
         \end{array}
     $$ 
\end{table}


The nominal errors as given by the cross-correlation are known to be
smaller than the true errors (e.g., Malumuth et al. \cite{mal92};
Bardelli et al. \cite{bar94}; Ellingson \& Yee \cite{ell94}; Quintana
et al. \cite{qui00}). Duplicate observations for the same galaxy
allowed us to estimate the true intrinsic errors in data of the same
quality taken with the same instrument (e.g. Barrena et
al. \cite{bar07a}, \cite{bar07b}).  Here we have double determinations
for eleven galaxies, thus we decided to apply the correction that had
already been applied in above studies. Hereafter we assume that true
errors are larger than nominal cross-correlation errors by a factor
of 1.3. For the eleven galaxies with two redshift estimates, we used
the weighted mean of the two measurements and the corresponding
errors. As for the radial velocities estimated through EMSAO we
assumed the largest between the nominal error and 100 \kss.  As for
the whole catalog, the median error in $cz$ is 52 \kss.

Our photometric observations were carried out in with the Wide Field
Camera (WFC), mounted at the prime focus of the 2.5m INT telescope. We
observed A545 in $g^{\prime}$, $r^{\prime}$ and $i^{\prime}$ Sloan-Gunn filters in photometric 
conditions and a seeing of $\sim$1.5\arcss.

The WFC consists of a four-CCD mosaic covering a
33\arcmm$\times$33\arcm field of view, with only a 20\% marginally
vignetted area. We took eight exposures of 600 s in $g^{\prime}$
filter, nine frames of 400 s exposure in $r^{\prime}$ filter and eight
exposures more of 400 s in $i^{\prime}$ filter. So a total of 4800 s
in $g^{\prime}$ filter, and 3600 s in $r^{\prime}$ and $i^{\prime}$
bands. The observations were developed making a dithering
pattern. This observing procedure allowed us to build a ``supersky''
frame that was used to correct our images for fringing patterns
(Gullixson \cite{gul92}). In addition, the dithering helped us to
clean cosmic rays and avoid the effects of gaps between the CCDs in
the final images.  Another problem associated with the wide field
frames is the distortion of the field. To match the photometry of
several filters, a good astrometric solution is needed to take into
account these distortions. Using the $imcoords$ IRAF tasks and taking
as a reference the USNO B1.0 catalog, we were able to find an accurate
astrometric solution (rms $\sim$0.4\arcss) across the full frame. The
photometric calibration was performed by observing standard Landolt
fields (Landolt \cite{lan92}) calibrated in the Sloan-Gunn system.

We finally identified galaxies in our $g^{\prime}$, $r^{\prime}$ and
$i^{\prime}$ images and measured their magnitudes with the SExtractor
package (Bertin \& Arnouts \cite{ber96}) and AUTOMAG procedure.
Objects were identified as imposing that they cover a certain minimum
area and have a number counts above a limiting threshold taking the
sky local background as a reference. The limiting size and flux were
16 pixels and 1.5 standard deviation of the sky counts,
respectively. The selected limiting size corresponds to an apparent
size of 1.3\arcss, which is about the minimum seeing size during the
observations. We have performed careful visual inspections of the
frames in order to deal with the best combination of the above
parameters that remove spurious objects from the catalogs.

In a few cases (e.g., close companion galaxies, galaxies close to
defects of the CCD) the standard SExtractor photometric procedure
failed. In these cases, we computed magnitudes by hand. This method
consisted of assuming a galaxy profile of a typical elliptical galaxy
and scaling it to the maximum observed value. The integration of this
profile provided an estimate of the magnitude. This method is similar
to PSF photometry, but assumes a galaxy profile, which is more
appropriate in this case.

As a final step, we estimated and corrected the Galactic extinction
$A_{g^{\prime}} \sim0.65$, $A_{r^{\prime}} \sim0.45$ and $A_{i^{\prime}}
\sim0.35$ using Schlegel et al. (\cite{sch98}) reddening and
extinction maps computed from IRAS and COBE/DIRBE data. We estimated
that our photometric sample is complete down to $g^{\prime}=23.6$
(24.5), $r^{\prime}=22.3$ (23.5) and $i^{\prime}=21.3$ (22.8) for
$S/N=5$ (3) within the observed field.

We assigned magnitudes to all galaxies of our spectroscopic catalog.
Table~\ref{catalogA545} lists the velocity catalog (see also
Fig.~\ref{figottico}): identification number of each galaxy, ID
(Col.~1); right ascension and declination, $\alpha$ and $\delta$
(J2000, Col.~2); $g^{\prime}$, $r^{\prime}$, and $i^{\prime}$
magnitudes (Cols.~3, 4 and 5); heliocentric radial velocity, ${\rm
  v}=cz_{\sun}$ (Col.~6) with error, $\Delta {\rm v}$ (Col.~7); and
code for membership (1/0 for members/non-members, respectively; Col.~8).

No evident dominant galaxy is present in the cluster, e.g. the
brightest galaxy in our catalog (ID.~2) is only 0.5 mag brighter in
the $r^{\prime}$ band then the second brightest galaxy (ID.~104) and
both are far from the cluster center as indicated by the peak of the
X-ray emission [R.A.=$05^{\mathrm{h}}32^{\mathrm{m}}25\dotsec6$,
  Dec.=$-11\degree 32\arcmm 37\arcs$ (J2000.0)] derived from our
analysis of the XMM-Newton data (see Sect. 4).  Other galaxies
brighter than 17 mag in the $r^{\prime}$ band are IDs.~27 and
89. Hereafter we refer to these galaxies as the BCG1, BCG2, BCG3, and
BCG4. These galaxies are $\gtrsim 0.4$ \h distant from the
  cluster center. As for the cluster central region, the most
luminous galaxy is ID.~64 (hereafter CBCG), which is located just
close to the peak of the X-ray emission and the location of the star
pile (see Figs.~\ref{figimage} and \ref{figottico}). For the CBCG
we measure a redshift concordant with that of the star pile (as
    obtained by S07 through Gemini-GMOS long-slit spectra).

Our spectroscopic catalog lists some radio galaxies observed in the
field of A545. Figure~2 of Bacchi et al. (\cite{bac03}) show four
discrete sources in the region of the radio halo: the optical
counterparts of S2 and S3 are the cluster members IDs.~27 and 29,
respectively. The radio source S4 is located in the crowed central
cluster region and its most likely counterpart is the luminous
CBCG. As for S1 there is no an obvious counterpart.  Figure~2 of
Bacchi et al.  also shows other three radio sources at the
border/outside the radio halo: the one at SE has as possible
counterpart the cluster member ID.~100; for the one at South we have
not acquired the redshift of the counterpart; the one at SW has no
counterpart in our optical image.

Another galaxy to be mentioned is ID.~57, which, according to S07, is
a very compact object. We find that it is a cluster member thus
discarding the possibility that it is a star projected onto the
cluster field.  Thus ID.~57 resembles the characteristics of a
M32-like dwarf object of which only a few representatives are known
(e.g., Chilingarian et al. \cite{chi07}; \cite{chi09}).  Moreover, we
find that its colors ($g^{\prime}$-$r^{\prime}$=0.91,
$r^{\prime}$-$i^{\prime}$=0.43) and spectrum resemble a typical early
type galaxy. Hereafter we indicate this galaxy as ``cE''.

\section{Analysis of the optical data}
\label{anal}

\subsection{Member selection}
\label{memb}

\begin{figure}
\centering
\resizebox{\hsize}{!}{\includegraphics{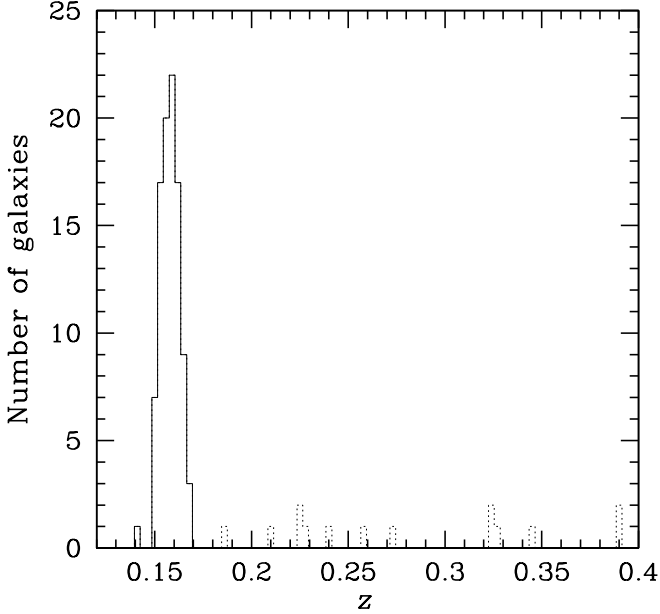}}
\caption
{Redshift galaxy distribution. The solid line histogram refers to the
  96 galaxies assigned to A545 according to the DEDICA
  reconstruction method.}
\label{fighisto}
\end{figure}

To select cluster members among the  110 galaxies with redshifts, we
follow a two-step procedure. We first perform the 1D
adaptive-kernel method (hereafter DEDICA, Pisani \cite{pis93} and
\cite{pis96}; see also Fadda et al. \cite{fad96}; Girardi et
al. \cite{gir96}). We search for significant peaks in the velocity
distribution at $>$99\% c.l.. This procedure detects A545 as a peak
at $z\sim0.157$ populated by 96 galaxies considered as candidate
cluster members (in the range $41\,964\leq {\rm v} \leq 50\,629$ \kss, see
Fig.~\ref{fighisto}). The 14 non-members are 1 and 14 foreground and
background galaxies, respectively.

\begin{figure}
\centering 
\resizebox{\hsize}{!}{\includegraphics{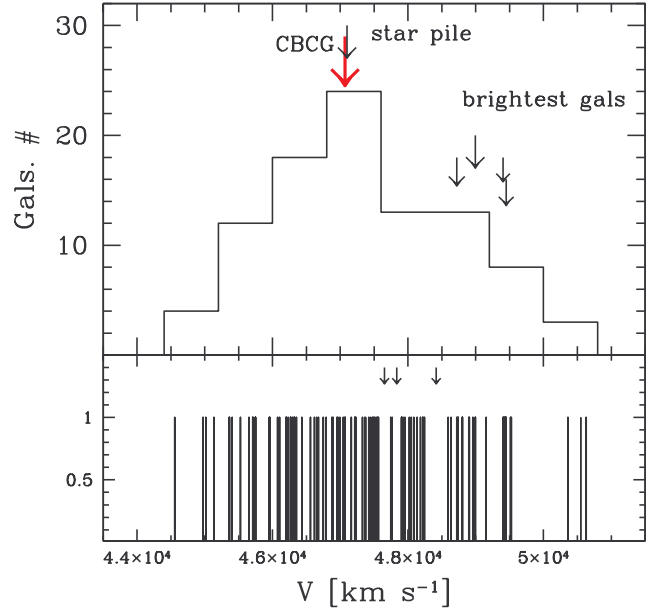}}
\caption
{The 95 galaxies assigned to the cluster.  {\em Upper panel}: Velocity
  distribution.  The arrows indicate the velocities of the four
  brightest galaxies, the velocity of the central bright galaxy CBCG 
  (bigger red arrow), and the velocity of the star pile.  
  {\em Lower panel}: Stripe density plot where the arrows
  indicate the positions of the significant gaps.}
\label{figstrip}
\end{figure}

All the galaxies assigned to the cluster peak are analyzed in the
second step, which uses the combination of position and velocity
information, i.e., the ``shifting gapper'' method by Fadda et
al. (\cite{fad96}).  This procedure rejects galaxies that are too far
in velocity from the main body of galaxies within a fixed bin that
shifts along the distance from the cluster center.  The procedure is
iterated until the number of cluster members converges to a stable
value.  Following Fadda et al. (\cite{fad96}), we use a gap of $1000$
\ks -- in the cluster rest-frame -- and a bin of 0.6 \hh, or large
enough to include 15 galaxies. As for the center of A545, we adopt the
position of the peak of the X-ray emission (see above).  The
``shifting gapper'' procedure rejects another one obvious interloper
very far from the main body ($>2000$ \kss) but that survived the first
step of our member selection procedure (see the red cross in the
Fig.~\ref{figprof}).  We obtain a sample of 95 fiducial members (see
Figs.~\ref{figstrip} and \ref{figprof}).

\subsection{Global cluster properties}
\label{glob}

By applying the biweight estimator to the 95 cluster members (Beers et
al. \cite{bee90}, ROSTAT software), we compute a mean cluster redshift
of $\left<z\right>=0.1580\pm$ 0.0004, i.e.
$\left<\rm{v}\right>=47373\pm$125 \kss.  We estimate the LOS
velocity dispersion, $\sigma_{\rm V}$, by using the biweight estimator
and applying the cosmological correction and the standard correction
for velocity errors (Danese et al. \cite{dan80}).  We obtain
$\sigma_{\rm V}=1220_{-68}^{+82}$ \kss, where errors are estimated
through a bootstrap technique.

\begin{figure}
\centering
\resizebox{\hsize}{!}{\includegraphics{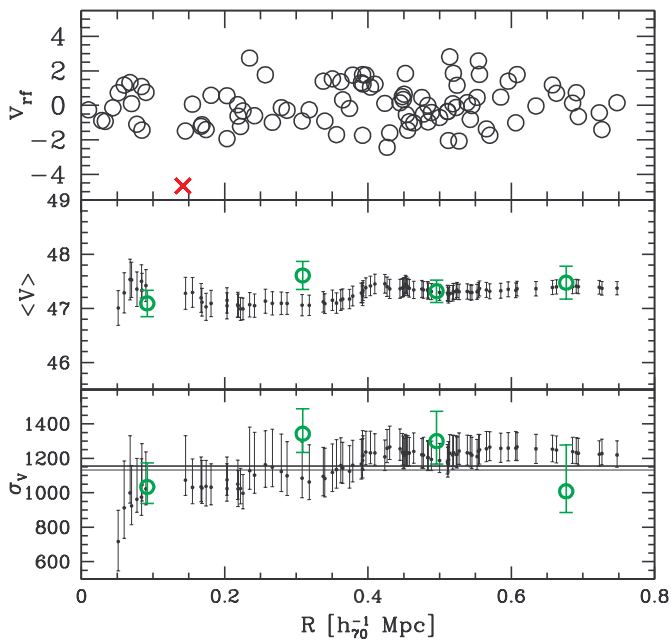}}
\caption
{{\em Top panel:} rest-frame velocity vs. projected distance from the
  cluster center of cluster members (circles). The red cross indicates
  the galaxy rejected by the ``shifting gapper'' selection procedure.
  {\em Middle and bottom panels:} differential (big circles) and
  integral (small points) profiles of mean velocity and LOS velocity
  dispersion, respectively.  For the differential profiles, we plot
  the values for four annuli from the center of the cluster, each of
  0.2 \h (large green symbols).  For the integral profiles, the mean
  and dispersion at a given (projected) radius from the
  cluster-center is estimated by considering all galaxies within that
  radius -- the first value computed on the five galaxies closest to
  the center. The error bands at the $68\%$ c.l. are also shown.  In
  the lower panel, the horizontal line represents the X-ray
  temperature with the respective errors transformed in
  $\sigma_{\rm V}$ assuming the density-energy equipartition between
  ICM and galaxies, i.e.  $\beta_{\rm spec}=1$ (see
  Sect.~\ref{disc}).}
\label{figprof}
\end{figure}

To evaluate the robustness of the $\sigma_{\rm V}$ estimate, we analyze
the velocity dispersion profile (Fig.~\ref{figprof}). The integral
profile rises out to $\sim0.1$ \h  and then flattens suggesting
that a robust value of $\sigma_{\rm V}$ is asymptotically reached in
the external cluster regions, as found for most nearby clusters (e.g.,
Fadda et al. \cite{fad96}; Girardi et al. \cite{gir96}).  

\subsection{Velocity distribution}
\label{velo}

We analyze the velocity distribution to search for possible deviations
from Gaussianity that might provide important signatures of complex
dynamics. For the following tests, the null hypothesis is that the
velocity distribution is a single Gaussian.

We estimate three shape estimators, i.e., the kurtosis, the skewness,
and the scaled tail index (see, e.g., Bird \& Beers \cite{bir93}).
According to the value of the normalized kurtosis ($KURT=0.627$), and
the scaled tail index ($STI=0.873$) the velocity distribution is
marginally differing from a Gaussian at the $90-95\%$ c.l. (see
Table~2 of Bird \& Beers \cite{bir93}).

We then investigate the presence of gaps in the velocity distribution.
We follow the weighted gap analysis presented by Beers et
al. (\cite{bee91}; \cite{bee92}; ROSTAT software).  We look for
normalized gaps larger than 2.25 since in random draws of a Gaussian
distribution they arise at most in about $3\%$ of the cases,
independent of the sample size (Wainer and Schacht~\cite{wai78}). We
detect three significant gaps (at the $99.4\%$, $97\%$, $99.95\%$
c.l.s), which divide the cluster into four groups of 58, 2, 11, and 24
galaxies from low to high velocities (hereafter GV1, GV2, GV3 and GV4,
see Fig.~\ref{figstrip}).  The CBCG is assigned to the GV1 group.  The
BCG1-2-3-4 are all assigned to the GV4 group.

Following Ashman et al. (\cite{ash94}), we also apply the Kaye's
mixture model (KMM) algorithm. This test does not find that a two or a
three or a four-groups partition would provide a significantly more
accurate description of the velocity distribution than a single
Gaussian.

\subsection{Analysis of the 2D galaxy distribution}
\label{photo}

\begin{figure}
\includegraphics[width=8cm]{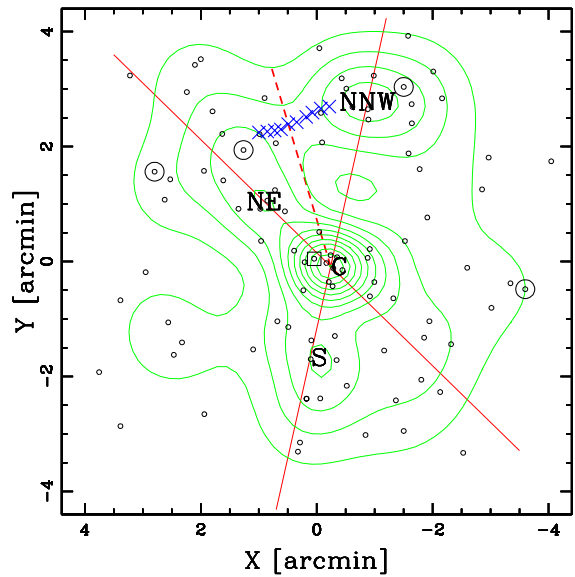}
\caption
{Spatial distribution on the sky and relative isodensity contour map
  of spectroscopic cluster members, obtained with the 2D DEDICA
  method.  The X-ray peak is taken as the cluster center.  The
  location of the four brightest galaxies are indicated by large
  circles.  The square indicates the central luminous galaxy CBCG. The
  four significant peaks are labelled. 
  The two (red) solid slides and the dashed line indicate the two likely 
  merging directions and their bisecting, respectively. Blue crosses 
  indicate, in a schematic way, the sharp discontinuity detected in the 
  X-ray surface brightness (see Sect. 4 for details).
  }
\label{figk2z}
\end{figure}

By applying the 2D adaptive-kernel method to the positions of A545
galaxy members, the most significant, dense peak is that coinciding
with the peak of the X-ray emission (``C'' label in
Fig.~\ref{figk2z}). Other three significant peaks are shown: one
located at North, slightly NNW); one located at NE; and one located at
South with respect to the position of the main one.  DEDICA
  assigns the CBCG and BCG1 to the C group, the BCG3 to the NNW group,
  and the BCG2 and BCG4 to the NE group. The CBCG, BCG3, and BCG4 are
  particularly close to their respective peaks (0.05, 0.11, and 0.16
  \h, respectively, see Fig.~\ref{figk2z}.  However, our
spectroscopic data do not cover the entire cluster field and are
affected by magnitude incompleteness.  To overcome these problems,
from our photometric catalog we use our photometric data sample which
covers a larger spatial region.

\begin{figure}
\centering
\includegraphics[width=8cm]{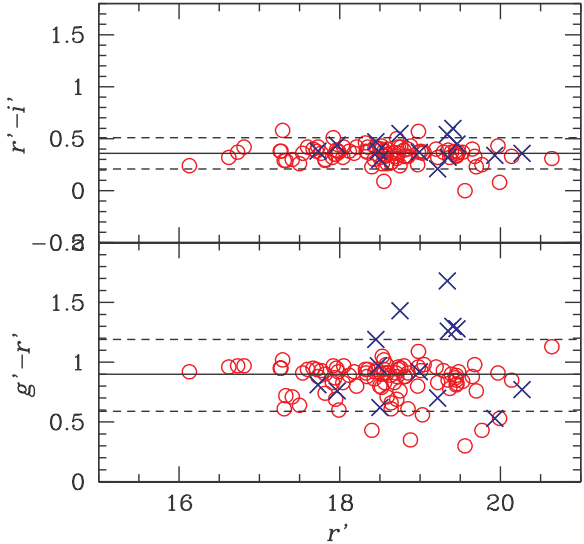}
\caption
{{\em Upper panel:} $r^{\prime}-i^{\prime}$ vs. $r^{\prime}$ diagram
  for galaxies with available spectroscopy.  Red circles and blue crosses
  indicate cluster members and non members. The solid line gives the
  median color determined on member galaxies; the dashed lines are
  drawn at $\pm$0.15 mag from this value.  {\em Lower panel:}
  $g^{\prime}-r^{\prime}$ vs. $r^{\prime}$ diagram for galaxies with
  available spectroscopy.  The dashed lines are drawn at $\pm$0.3 mag
  from the median value determined on member galaxies.  }
\label{figcm}
\end{figure}

In our photometric catalog we select likely members on the base of both
($r^{\prime}$--$i^{\prime}$) and ($g^{\prime}$--$r^{\prime}$) colors.
Goto et al. (\cite{got02}) showed that there is a small tilt in the
color-magnitude relations ($r^{\prime}$--$i^{\prime}$)
vs. $r^{\prime}$ and ($g^{\prime}$--$r^{\prime}$) vs. $r^{\prime}$ and
that the scatter in the latter relation is roughly the double than the
scatter in the first one (0.040 and 0.081 mags, respectively).
Out of our photometric catalog we consider as ``likely cluster members''
those objects having ($r^{\prime}$--$i^{\prime}$) and
($g^{\prime}$--$r^{\prime}$) lying within $\pm$0.15 and $\pm$0.3 from
the median values of $r^{\prime}$--$i^{\prime}$=0.36 and
$g^{\prime}$--$r^{\prime}$=0.89 colors of the spectroscopically
cluster members, i.e., we reject those objects having colors
  $\gtrsim 3$ sigma far from the red sequence (see Fig.~\ref{figcm}).

\begin{figure}
\includegraphics[width=8cm]{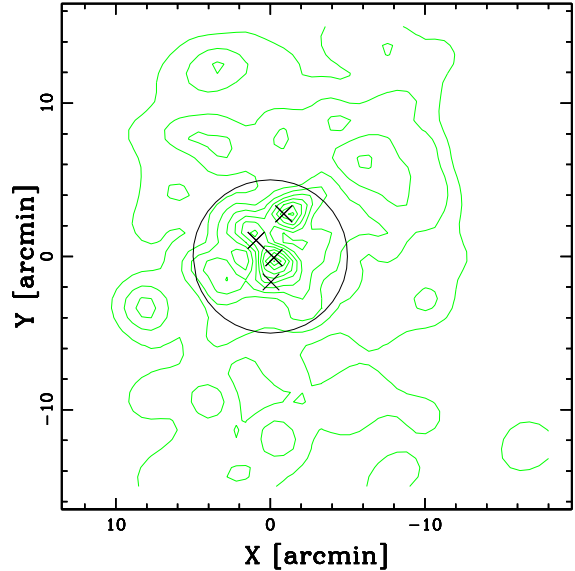}
\caption
{Spatial distribution on the sky and relative isodensity contour map
  of likely cluster members with $r^{\prime}\le 20$, obtained with the
  2D DEDICA method. The large faint circle indicate the region of
  5\arcm-radius which is sampled by the spectroscopic data.  The
  crosses indicate the position of the peaks in the distribution of
  the spectroscopic members.  
}
\label{figk2}
\end{figure}

Figure~\ref{figk2} shows the contour map for the 902 galaxies
  selected on the base of the cluster red sequence having
$r^{\prime}\le 20$.  The most significant peaks coincide with the peak
in the cluster core, the NNW peak, and the NE peak detected in the 2D
distribution of spectroscopic members.  Similar results are found for
the 1304 objects with $r^{\prime}\le 21$.

As for the results with $r^{\prime}\le 20$, Table~\ref{tabdedica2d}
lists the number of assigned members, $N_{\rm S}$ (Col.~2); the
peak position (Col.~3); the density (relative to the densest
peak), $\rho_{\rm S}$ (Col.~4); the value of $\chi^2$ for
  each peak, $\chi ^2_{\rm S}$ (Col.~5); the distance of the
  closest BCG from the peak position with the name of the respective
  BCG (Col.~6).  Ramella et al. (\cite{ram07}) tested the 2D DEDICA
procedure on Monte-Carlo simulations reproducing galaxy clusters. They
show that the physical significance, i.e. the significance which takes
into account the noise fluctuations, associated to the subclusters
depends on the statistical significance of the subcluster (recovered
from the $\chi ^2_{\rm S}$ value) and can be computed using
simulations.  Comparing the values for the three peaks in A545 with
the Fig.~2 by Ramella et al. (\cite{ram07}, obtained from simulations
of 900 data points) we conclude that these substructures are less than
7\% of probability to be statistical artifacts. In our analysis other
peaks have $\chi ^2_{\rm S}<17$ and thus a probability larger than
10\% to be artifacts.

\begin{table}
        \caption[]{Substructure from the photometric sample with 
$r^{\prime}\le 20$ found with DEDICA.}
         \label{tabdedica2d}
            $$
         \begin{array}{l r c c c l}
            \hline
            \noalign{\smallskip}
            \hline
            \noalign{\smallskip}
\mathrm{Subclump} & N_{\rm S} & \alpha({\rm J}2000),\,\delta({\rm J}2000)&\rho_{\rm S}&\chi^2_{\rm S}& dist\mathrm{(BCG)}\\
& & \mathrm{h:m:s,\degree:\arcmm:\arcs}&&&h_{70}^{-1} \mathrm{Mpc}\\
         \hline
         \noalign{\smallskip}
\mathrm{2D-C}   & 73&05\ 32\ 24.6 , -11\ 32\ 52&1.00&31&0.07\mathrm{(CBCG)}\\
\mathrm{2D-NNW} & 58&05\ 32\ 20.4 , -11\ 29\ 49&0.85&25&0.08\mathrm{(BCG4)}\\
\mathrm{2D-NE}  & 36&05\ 32\ 29.8 , -11\ 31\ 07&0.66&20&0.05\mathrm{(BCG3)}\\
              \noalign{\smallskip}
            \hline
            \noalign{\smallskip}
            \hline
         \end{array}
$$
         \end{table}

\begin{figure}
\includegraphics[width=8cm]{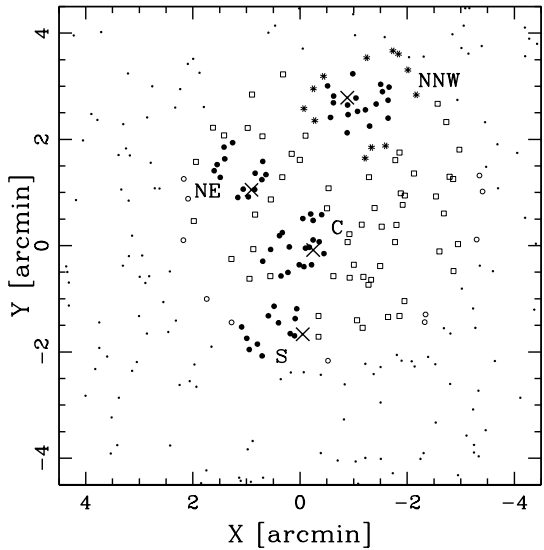}
\caption{Galaxies belonging to structures as detected by the Voronoi
  Tessellation and Percolation technique. The algorithm is run on the
  sample of likely members with $r^{\prime}\le$21 (see text). Open
  circles, open squares, asterisks and solid circles indicate galaxies
  in structures at the 90\%, 95\%, 98\% and 99\% c.ls.,
  respectively. The C, NE, NNW and S subclumps are indicated. As a
  comparison, big crosses indicate the positions of the peaks in the
  distribution of the spectroscopic members.}
\label{figvtp}
\end{figure}

To further probe the robustness of these detections, we also apply the
Voronoi Tessellation and Percolation (VTP) technique (e.g. Ramella et
al. \cite{ram01}; Barrena et al. \cite{bar05}). This technique is
non-parametric and does not smooth the data. As a consequence, it
identifies galaxy structures irrespective of their shapes.  The result
of the application of VTP on the sample of 1304 likely members with
$r^{\prime}\le 21$ is shown in Fig.~\ref{figvtp}. VTP is run four
times adopting four detection thresholds: galaxies identified as
belonging to structures at 90\%, 95\%, 98\% and 99\% c.ls. are shown
as open circles, open squares, asterisks and solid circles
respectively. VTP confirms the central peak and the NE and NNW ones
found by DEDICA in the photometric sample. However, in this case VTP
also finds a significant peak located 2\arcm South of the cluster
center, roughly coincident with the peak detected by DEDICA in the
distribution of spectroscopic members. As for the detection of
  the southern peak, the difference between DEDICA and VTP applied to
  the same sample of photometric members can be explained by the fact
  that DEDICA, like any kernel-based technique in general, is more
  sensitive to symmetric structures than asymmetric ones. The result
  is that the southern overdensity, which is very close and less dense
  than the central one, causes only a southern elongation of the
  iso-density contours in the DEDICA central peak (see
  Fig.~\ref{figk2}, also more evident in the corresponding figure for
  $r^{\prime}\le 21$). On the contrary, the non-parametric nature of
  VTP favors an easier separation of the central and southern peaks
  (cf.  Fig.~\ref{figvtp}).

\subsection{3D-analysis}
\label{3d}

Here we use different approaches to analyze the structure of
A545 combining position and velocity information.  

To search for a possible physical meaning of the three subclusters
determined by the two weighted gaps, we compare two by two the spatial
galaxy distributions of GV1, GV3, and GV4; GV2 is not considered
  since is only formed by two galaxies.  We find no difference
according to the the 2D Kolmogorov-Smirnov test (Fasano \&
  Franceschini \cite{fas87}).

We analyze the presence of a velocity gradient performing a multiple
linear regression fit to the observed velocities with respect to the
galaxy positions in the plane of the sky (e.g., den Hartog, R., \&
Katgert \cite{den96}; Girardi et al. \cite{gir96}; Boschin et
al. \cite{bos04}). To assess the significance of this velocity
gradient we perform 1000 Monte Carlo simulations by randomly shuffling
the galaxy velocities and for each simulation we determine the
coefficient of multiple determination ($RC^2$, see e.g., NAG Fortran
Workstation Handbook \cite{nag86}).  We define the significance of the
velocity gradient as the fraction of times in which the $RC^2$ of the
simulated data is smaller than the observed $RC^2$. We find that the
velocity gradient is not significant.

In order to check for the presence of substructure, we combine
  velocity and position information by computing the
  $\Delta$-statistics devised by Dressler \& Schectman (\cite{dre88},
  hereafter DS-test), which is recommended by Pinkney et al.
  (\cite{pin96}) as the most sensitive 3D test.  For each galaxy, the
  deviation $\delta$ is defined as $\mathrm{\delta^2 =
    [(N_{\rm{nn}}+1)/\sigma_{\rm{V}}^2][(\overline V_l - \overline
      V)^2+(\sigma_{\rm{V},l} - \sigma_{\rm{V}})^2]}$, where the
  subscript ``$l$'' denotes the local quantities computed over the
  $N_{\rm{nn}}=10$ neighbors of the galaxy.  $\Delta$ is the sum of
  the $\delta$ of the individual $N$ galaxies and gives the cumulative
  deviation of the local kinematical parameters (mean velocity and
  velocity dispersion) from the global cluster parameters.  We find
  $\Delta=108$.  The significance of substructure is again checked by
  running 1000 Monte Carlo simulations, randomly shuffling the galaxy
  velocities.  We find that 336 simulated clusters show a value of the
  $\Delta$ parameter larger than that of the real cluster leading to
  non significant presence of substructure.

Following Pinkney et al. (\cite{pin96}; see also Ferrari et
al. \cite{fer03}), we apply other two classical 3D tests: the
$\epsilon$-test (Bird, \cite{bir93}) based on the projected mass
estimator and the centroid shift or $\alpha$-test (West \& Bothun
\cite{wes90}). The details of these tests can be found in the above
papers: we only point out that we consider ten as the number of the
nearest neighbors for each galaxy and we use the above Monte Carlo
simulations to compute the substructure significance.  In both cases
we do not find evidence for significant substructure.

We also consider three kinematical estimators alternative to the
$\delta$ parameter of the DS-test, i.e. we consider separately the
contributes of the local mean $\mathrm{\delta_V^2 =
  [(N_{\rm{nn}}+1)/\sigma_{\rm{V}}^2](\overline V_l - \overline
  V)^2}]$, and dispersion $\mathrm{\delta_s^2 =
  [(N_{\rm{nn}}+1)/\sigma_{\rm{V}}^2](\sigma_{\rm{V},l} -
  \sigma_{\rm{V}})^2}]$ (see, e.g.  Girardi et al. \cite{gir97},
  Ferrari et al. \cite{fer03}).  For both of the above estimators, we
  do not find evidence for significant substructure.

\section{X-ray morphological and spectral analysis}
\label{xray}

We retrieved from the XMM-Newton archive the available observation of
A545 (Obs. ID.~0304750101), which has a nominal exposure time of 36.4
ks.  We reprocessed the Observation Data Files (ODF) using the Science
Analysis System (SAS) version 9.0.  After the production of the
calibrated event lists for the EPIC MOS1, MOS2 and {\it pn}
observations with the {\it emchain} and {\it epchain} tasks, we tried
to perform a soft proton cleaning.  Unfortunately, by screening the
light curves, produced in 100 s bins in the 10--12 keV band, we
discovered that this observation is highly affected by soft protons
flares.  The contamination is so severe that we could not apply the
standard count rate thresholds for the MOS and {\it pn} (i.e., 0.15
cts/s and 0.35 cts/s, respectively). We decided to apply higher
thresholds, namely 1.6 cts s$^{-1}$ for MOS1 and MOS2 and 4.5 cts
s$^{-1}$ for {\it pn}, and then use these least contaminated data to
extract global X-ray properties in the highest surface brightness
cluster regions only, where the ICM photons clearly dominate over the
soft proton background.  We also filtered event files according to
FLAG (FLAG=0) and PATTERN (PATTERN$\leq 12$ for MOS and PATTERN=0
for {\it pn}) criteria.  The resulting effective exposure time of the
observation is 16 ks for the MOS and 9 ks for the {\it pn}.

\subsection {Surface brightness analysis}
\label{xs}

The EPIC (i.e. with all the three detectors summed together) surface
brightness contours of A545 in the 0.4--2 keV band, overplotted to the
INT $r^{\prime}$-band optical image, are shown in Fig.~\ref{figimage}.
We produce the X-ray image following the procedure described in 
Rossetti et al. (\cite{ros07}, see their Fig.~6).

The analysis of the surface brightness map reveals that the cluster is
strongly elongated NNW-SSE.  Following the procedure of Hashimoto et
al. (\cite{has07}), we compute the ellipticity of the isophotes. We find
that the cluster is indeed strongly elongated, with an ellipticity
$\epsilon=1-b/a=0.37$, where $a$ and $b$ are the major and minor axes
of the isophotes, respectively. 

To produce the surface-brightness profile (see Eckert et
al. \cite{eck11}), we extract MOS1 and MOS2 count images in a broad
band (0.4-8 keV), to include as many photons as possible, and produce
the corresponding exposure maps using the SAS tool {\it eexpmap} to
correct for vignetting effects. Point sources are excised from
  the image using a local background method. Namely, we excise regions
  of the image where the observed count rate is at least 5 sigma above
  the local background. We then extract the surface-brightness
profile of the cluster centered on the emission centroid in elliptical
annuli, with bins of width 7\arcss.  More specifically, we select
regions with equal elliptical distance to the center, where the
distance is given by $R=\sqrt{\frac{a^2}{b^2}x_1^2+x_2^2}$, $x_1$ and
$x_2$ being the distances along the major and minor axes. The profile
is then fitted by a standard $\beta$ model (Cavaliere \& Fusco-Femiano
\cite{cav76}),

\begin{equation}
SB (R)= S_0[1+(R/R_{\rm c})^2]^{-3\beta+1/2}+B,
\end{equation}

where $R_{\rm c}$ is the core radius along the major axis, $\beta$
gives the slope of the outer density profile, and $B$ is a constant to
determine the background level. The surface brightness profile
extracted in elliptical annuli is rather well represented
($\chi^2_{\rm red}=112.5/73$) by a beta profile with
$\beta=0.87_{-0.02}^{+0.03}$ and a core radius $R_{\rm
  c}=2.03\pm0.06\arcm$ ($333\pm10$ \kpc) along the major axis, which
corresponds to a core radius of $1.28\pm0.04\arcm$ ($210\pm7$ kpc)
along the minor axis (this profile is plotted as a blue line in
  Fig. 11).

\begin{figure}
\centering 
\includegraphics[width=8cm]{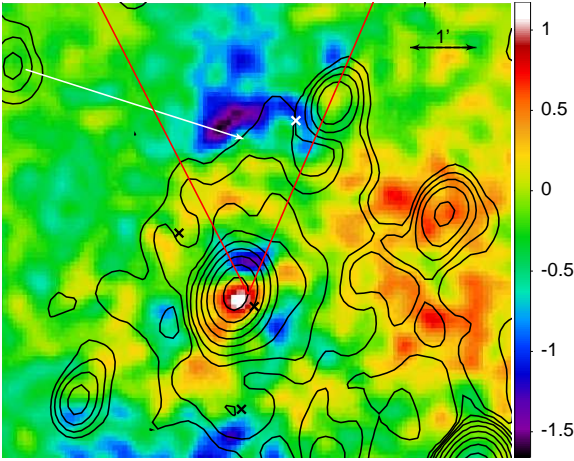}
\caption{ Residuals of the X-ray surface brightness map after
  subtraction of the best fitting elliptical beta model (the colors
  correspond to the deviations in sigma to the surface brightness
  profile for each pixel). The black contours show the extended radio
  emission (Bacchi et al. \cite{bac03}). The arrow indicates the
  position of the North surface brightness jump. The two red lines
    show the 70-120 deg sector used to produce the profile plotted in
    Fig.~\ref{northfeaturenew}. Crosses show the positions of
    the peaks in the distribution of the optical spectroscopic members
    as in Figs.~\ref{figk2}, and  ~\ref{figvtp}.}
\label{a545deviations}
\end{figure}

\begin{figure}
\centering 
\includegraphics[width=7.9cm,angle=-90]{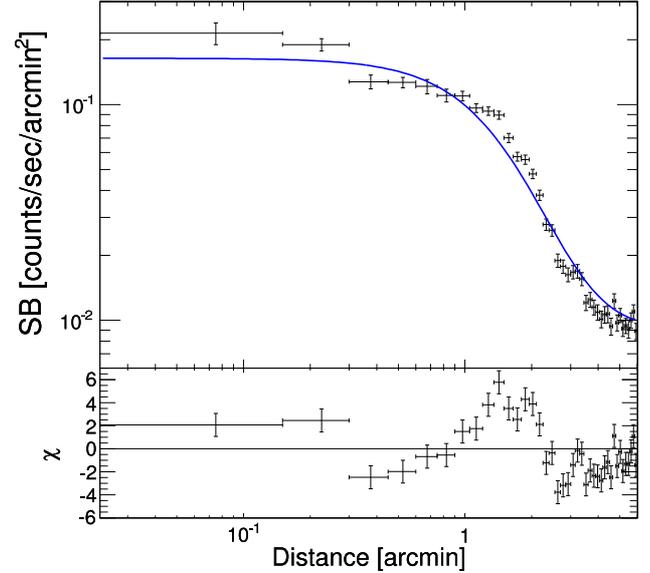}
\caption{ XMM-Newton surface brightness profile in direction of the North
    feature (from a sector with position angles 70-120 deg). The blue
    line is the best-fitting single beta model for the total
    profile. At $\sim 2.7\arcm$ from the center there is a clear
    surface brightness jump.}
\label{northfeaturenew}
\end{figure}

We look for substructures in A545 by removing the best-fit elliptical
beta model from the surface brightness image.  The residuals, in units
of standard deviation per pixel, are shown in
Fig.~\ref{a545deviations}, with radio contours overlaid.  At $\sim
2.7\arcm$ ($\sim 440$ kpc) North of the X-ray centroid, we note a
significant depletion (or ``jump'') with respect to the average
profile, that could be a signature of a cold front or of a shock (see
also Fig.~\ref{a545deviations}).  To investigate this possibility, we
extract the surface brightness profile in a sector with position
angles 70--120 deg (with respect to the R.A. axis; the sector is shown
in Fig.~\ref{a545deviations}). In Fig.~\ref{northfeaturenew} we plot
the surface brightness profile in this particular sector, compared to
the average profile of the whole cluster, described above. To
characterize more quantitatively the jump we compare the slope of the
mean profile with the slope found in profile of the 70-120 deg sector
by modeling both profiles in the radial range $100\arcss-200\arcs$
with a simple power law model. We find that the difference of the
slope estimated in the 70-120 deg sector with respect to the slope
from the mean profile is 0.77$\pm$ 0.11.  


Other interesting features noticeable from the residuals shown in
Fig.~\ref{a545deviations} are the asymmetry along the E-W axis, the
W side showing an excess of emission compared to the E, and a surface
brightness decrement $\sim30\arcss$ North of the emission
peak.  While the E-W asymmetry might be worthwhile to be investigated
with new XMM-Newton data of better quality, the decrement right N of
the X-ray emission peak could be due to an imperfect subtraction of a
point-like source (at R.A.$=05^{\rm h} 32^{\rm m} 24\dotsec91$ and Dec.$=-11\degree 31\arcmm 39\dotarcs97$) and it would need the higher
resolution of a {\it Chandra} observation to be investigated in
details. 

\subsection {Spectral analysis}
\label{xt}

We apply the background subtraction using blank-sky fields for EPIC
MOS and {\it pn} that were produced by Leccardi et al. (\cite{lec08}, see 
Appendix B in their paper) by analyzing a large number of
observations for a total exposure time of $\sim 700$ ks for MOS and
$\sim 500$ ks for {\it pn}.  To take into account temporal variations
of the background, more than possible in this highly flared
observation, we also perform a background rescaling by estimating the
background intensity in this observation from a spectrum extracted in
an external annulus between $10^\prime$ and $12^\prime$ centered on
the emission peak in the $10-12$ keV band (to avoid possible extended
cluster emission residuals in this region). We then rescale the
blank-sky fields background to the local value.

Using the XSPEC package (version 11.3.2, Arnaud et al. \cite{arn96}),
we analyze the EPIC spectrum of a circular region of 310 kpc radius
centered on the X-ray peak (we exclude the pointlike source North of
the center). We model the spectrum with a one temperature thermal
model with the plasma in collisional ionization equilibrium ({\it
  mekal} model in the XSPEC nomenclature), multiplied by the Galactic
hydrogen column density, $N_{\rm H}$ = 1.02E+21 cm$^{-2}$, determined
by HI surveys (Kalberla et al. \cite{kal05}) through the {\it wabs}
absorption model.  The redshift is fixed to $z = 0.158$. The Fe
abundance is measured relative to the solar photospheric values of
Anders \& Grevesse (\cite{and89}), where Fe $= 4.68\times 10^{-5}$.
We find that the ICM has a global temperature of $8.08\pm
0.32$ keV and Fe abundance of $0.30\pm 0.05$ (solar
units).

We apply the same analysis to the spectrum extracted from a circular
region of radius 0.05$r_{\rm 180}=0.6\arcm$ (105 kpc) centered on
the X-ray emission peak, which is the dimension of a typical cool core
region (e.g., Leccardi et al. \cite{lec10}).  The resulting
temperature does not show any significant decrement within the errors
($8.42\pm0.48$ keV), as expected for a non cool core cluster.

We also tried to provide a more detailed description of the
temperature and abundance structure of A545. For example,
investigating differences of surface brightness, temperature and
abundance in regions coincident with the optical subclumps, or
producing thermodynamical maps (e.g. temperature, pressure, entropy)
using the procedure described in Rossetti et
al. (\cite{ros07}). However, the significant residual soft protons
contamination -- the background of the "cleaned " dataset is a factor
of 4 larger than that of a normal observation -- prevents us from
providing this more detailed description of A545.

\section{Discussion and conclusions}
\label{disc}

We have analyzed for the fist time the internal dynamics of A545. 
We find $\left<z\right>=0.1580\pm$0.0004, high values of the velocity 
dispersion $\sigma_{\rm V}=1220_{-68}^{+82}$ \ks and X-ray temperature $kT_{\rm
  X}=8.08\pm0.32$ keV. These results are comparable to the highest values found in
typical clusters (see Mushotzky \& Scharf \cite{mus97}; Girardi \&
Mezzetti \cite{gir01}).  
Our estimates of $\sigma_{\rm V}$ and $T_{\rm X}$ are fully consistent
when assuming the equipartition of energy density between ICM and
galaxies. We obtain $\beta_{\rm spec} =1.12^{+0.15}_{-0.12}$ to be
compared with $\beta_{\rm spec}=1$, where $\beta_{\rm
  spec}=\sigma_{\rm V}^2/(kT_{\rm X}/\mu m_{\rm p})$ with $\mu=0.58$ the mean
molecular weight and $m_{\rm p}$ the proton mass (see also
Fig.~\ref{figprof}).

In the framework of usual assumptions (cluster sphericity, dynamical
equilibrium, coincidence of the galaxy and mass distributions), one
can compute virial global quantities. Following the prescriptions of
Girardi \& Mezzetti (\cite{gir01}, Eq.~1 with the scaling of $H(z)$), we assume for the radius of the
quasi-virialized region is $R_{\rm vir}=0.17\times \sigma_{\rm
  V}/H(z) = 2.74$ \hh, see
also Eq.~ 8 of Carlberg et al. (\cite{car97}) for $R_{200}$.  We
compute the virial mass (Limber \& Mathews \cite{lim60}; see also,
e.g., Girardi et al. \cite{gir98}):

\begin{equation}
M=3\pi/2 \cdot \sigma_{\rm V}^2 R_{\rm PV}/G-{\rm SPT},
\end{equation}

\noindent where SPT is the surface pressure term correction (The \&
White \cite{the86}) and $R_{\rm PV}$ is a projected radius (equal to
two times the projected harmonic radius).

The estimate of $\sigma_{\rm V}$ is robust when computed within a
large cluster region (see Fig.~\ref{figprof}).  The value of $R_{\rm
  PV}$ depends on the size of the sampled region and possibly on the
quality of the spatial sampling (e.g., whether the cluster is
uniformly sampled or not).  Since in A545 we sample only a fraction
of $R_{\rm vir}$, we have to use an alternative estimate of $R_{\rm
  PV}$ on the basis of the knowledge of the galaxy
distribution. Following Girardi et al. (\cite{gir98}; see also Girardi
\& Mezzetti \cite{gir01}), we assume a King-like distribution with
parameters typical of nearby/medium-redshift clusters: a core radius
$R_{\rm c}=1/20\times R_{\rm vir}$ and a slope-parameter $\beta_{\rm
  fit,gal}=0.8$, i.e. the volume galaxy density at large radii goes as
$r^{-3 \beta_{\rm fit,gal}}=r^{-2.4}$. We obtain $R_{\rm PV}(<{\rm
  R}_{\rm vir})=2.04$ \hh, where a $25\%$ error is expected (Girardi
et al. \cite{gir98}, see also the approximation given by their Eq.~13
when $A=R_{\rm vir}$). The value of SPT depends strongly on the radial
component of the velocity dispersion at the radius of the sampled
region and could be obtained by analyzing the (differential) velocity
dispersion profile, although this procedure would require several
hundreds of galaxies. We decide to assume a $20\%$ SPT correction as
obtained in the literature by combining data on many clusters sampled
out to about $R_{\rm vir}$ (Carlberg et al. \cite{car97}; Girardi et
al. \cite{gir98}).  We compute $M_{\mathrm{one-cluster}}(<R_{\rm vir}=2.74
\hhh)=2.7_{-0.7}^{+0.8}$ \mquii.

\subsection{Internal structure}
\label{disc1}

The appearance of this cluster is well far from being that of
a relaxed cluster. Evidence of this comes from both optical and X-ray
analyses.

The position of the highest peak of the 2D galaxy distribution
coincides with the position of the peak in the X-ray distribution, but
other two, maybe three, galaxy clumps are detected. One is located
$\sim 3$\arcm at North, slightly NNW (close to the BCG3), and seems a
well separated unit from the cluster core. The other is located $\sim
2$\arcm at NE, with isodensity contours elongated toward to the
BCG4. The possible, fourth peak (detected by the VTP analysis) lies at
South. Very interestingly, the southern clump, if real, well
delineates with the NNW clump a possible direction for the cluster
accretion (see Figs.~\ref{figk2z} and \ref{figvtp}). Another possible
direction for the cluster accretion is the NE-SW direction as
also suggested by the apparent SW extension of the radio halo
at large scale (see Fig.~2 of Bacchi et al. \cite{bac03}) and
the small scale features concerning the star pile (see Fig.~1 of
  S07, see also the insect figure in Fig.~\ref{figottico} and
Sect.~\ref{disc2}).

The integral velocity dispersion profile increases from the central
cluster regions out to 0.1 \h (see Fig.~\ref{figprof}, bottom panel).
Although an increase in the velocity-dispersion profile in cluster
central regions might be due to dynamical friction and galaxy merging
(e.g., Menci \& Fusco-Femiano \cite{men96}; Girardi et
al. \cite{gir98}; Biviano \& Katgert \cite{biv04}), in the case of the
substructured cluster A545 it is likely induced by the contamination
of the galaxies of a secondary clump, i.e. to the mix of galaxies
  of two systems having different mean velocities (e.g., Girardi et
al. \cite{gir96}; Barrena al. \cite{bar07b}). The increase of
    $\sigma_{\rm V}$ due to the contamination starts well before the
    inclusion of (the center of) the companion system, e.g.  leading to
    ``uncontaminated'' regions of 0.3-0.4 \hc for rich clusters having
    a companion at 1 \hc (see A3391 and A3395 in Fig.~1 of Girardi et
    al. \cite{gir96}). Assuming a similar scenario for A545, the
    increase of the $\sigma_{\rm V}$ profile in the central region
    shown by Fig.~\ref{figprof} is due to the via via increasing
    contamination of the close subclumps (note that the NE clump is
    only $\sim 0.3$ \h far from the C clump). We take the value of the
    velocity dispersion $\sigma_{\rm V}\sim 700$ \kss, based on the
    five most central galaxies within 0.05 \h (i.e. the first left
    point plotted in Fig.~\ref{figprof}, bottom panel), as the most
    representative value of the velocity dispersion of the
    ``uncontaminated'' main C subcluster.  Following the above virial
    mass estimation procedure or, simpler, applying the usual
    scaling-laws where $R_{\rm vir} \propto \sigma_{\rm V}$ and
    $M(<R_{\rm vir}) \propto \sigma_{\rm V}^3$, we use the value of
    $\sigma_{\rm V}\sim 700$ \ks to estimate the virial mass of C
    subcluster as $M_{C}(<R_{\rm vir}\sim1.6 \hhh)\sim 0.5$ \mquii.

We also analyze the $\sigma_{\rm V}$-profiles corresponding to the NNW
and NE galaxy clumps, but there is no trace of a increase of the
$\sigma_{\rm V}$-profiles, rather the central values of $\sigma_{\rm
  V}$ are very high ($>1500$ \kss) suggesting that these groups are
already contaminated in their central regions and cannot be detected
as separate units.  This result prevents us to attempt an alternative
estimate of the whole A545 complex as the combination of the three
subclusters directly from the kinematics of galaxies.

We can attempt an alternative procedure using galaxy number as a proxy
for the mass. According to our analysis of the 2D substructure
(Table~\ref{tabdedica2d}), we can compute the mass of the cluster from
the three main clumps as $M_{\mathrm{sum-subclusters}}(R<1.6
\hhh)=M_{\rm C}\cdot N_{\rm C+NNW+NE}/N_{\rm C}= 0.5\cdot 167/73\sim
1.1$ \mquii.  Since we do not consider the likely southern subcluster,
this value can be taken as a lower limit for the mass estimate.  To
rescale the value of $M_{\mathrm{one-cluster}}$we assume that the
system is described by a King-like mass distribution (see above) or,
alternatively, a NFW profile where the mass-dependent concentration
parameter $c$ is taken from Navarro et al. (\cite{nav97}) and rescaled
by the factor $1+z$ (Bullock et al. \cite{bul01}; Dolag et
al. \cite{dol04}). We obtain $M_{\mathrm{one-cluster}}(R<1.6 \hhh)\sim
(1.7$--$1.8)$ \mquii.  Therefore, we estimate the range for the mass of
A545 as $M(R<1.6 \hhh)=(1.1$--$1.8)$ \mquii. Note that, assuming an
isothermal $\beta$ model for the ICM (e.g., eq. 6 of Henry et
al. \cite{hen93}) and using the global ICM temperature and the fitted
parameters of the $\beta$ model of Sect.~\ref{xray}, we obtain from
X-ray data a mass $M_{\mathrm X}(R<1.6 \hhh)\sim 1.3$ \mquii, falling
in the estimated mass range.  According to the usual scalings, the
above mass range leads to these global virial quantities for A545:
$\sigma_{\rm V}\sim (900$--$1200)$ \ks and $M_{\mathrm vir}(<R_{\rm
  vir}=2$--$2.7 \hhh)=(1.5$--$2.7)$ \mquii.

As for the view angle of the merger axis, we have three indications
that the merger is mostly occurring in the plane of the sky: i) the
successful detection of substructure mostly through the 2D analysis
suggests that the intervening subclusters are very close in the
velocity space (see Pinkney et al. \cite{pin96}); ii) the mean
velocity profile of the C galaxy clump shows an increase in the central
region, likely due to the contamination of close clumps, but
quite modest $\sim 500$ \ks (Fig.~\ref{figprof}, central panel); iii)
the likely presence of a shock in the ICM (see below) since a shock is
more easily discernible in an X-ray image when the cluster merger
occurs in the plane of the sky (e.g., Markevitch \cite{mar10}).

Even in the case of very unrelaxed clusters, the brightest galaxies
seem to trace the galaxy or dark matter subclumps both in two
dimensions (Beers \& Geller \cite{bee83}) and in 3D space (e.g. Abell
520, Girardi et al. \cite{gir08}).  In A545 the brightest galaxies
BCG1, BCG2, BCG3, and BCG4 avoid the cluster core and are far from the
mean cluster velocity, but the BCG3 and BCG4 seem to be related with two
peaks in the galaxy distribution (NNW and NE clumps). Instead, the
brightest galaxy in the cluster core (CBCG), although is more than 1
mag fainter then the BCG1, is related to the main density peak and is at
rest with respect to the whole cluster rest frame. 

The indications that we derive from the analysis of the X-ray
surface brightness distribution (namely: the deviations from spherical
symmetry, the northern discontinuity, the western excess)
clearly provide evidence of a disturbed dynamical phase, too.  A545 is
indeed strongly elongated, with an ellipticity $\epsilon=0.37$.  From
the sample of Hashimoto et al. (\cite{has07}), we see that such a
value would classify A545 as one of the most elliptical objects. Since
the authors note a correlation between the ellipticity and the
dynamical state, this supports a scenario where the cluster is
dynamically active. Moreover, the elongation is mostly in the 
  NNW-SSE direction indicating the same direction of the NNW galaxy
clump with respect to cluster center.  Another hint on the unrelaxed
state of A545 comes also from the clear twisting of the isophotes
observable from the center towards the more external regions (see
Fig.~\ref{figimage}).  As for the western X-ray excess, very
noticeably the 2D galaxy distribution reveals an excess in the western
regions, too. In fact, looking at Fig.~\ref{figvtp}, we note
  that the distribution of galaxies assigned to the cluster with a
  probability of $\gtrsim 95\%$ ($\gtrsim 90\%$) do not follow that of
  galaxies with highest probabilities, neither is symmetric
  around the cluster center.  The result is that, out of galaxies with
  a probability of $\gtrsim 95\%$ ($\gtrsim 90\%$), we find 13 (18)
  galaxies 2\arcm west from the cluster center (where we start to detect
 the X-ray excess) and no (3) galaxies 2\arcm east from
  the cluster center.

The most interesting feature in the X-ray surface brightness is the
northern sharp discontinuity ($\sim 2.7\arcm$ with respect to the
X-ray centroid), which is a signature of a shock or a cold front
(e.g., Markevitch \cite{mar10} and refs. therein).  Interestingly, the
shock possibility is supported by the fact that at the location of the
surface brightness jump the giant radio halo also shows an edge (see
Fig.~\ref{a545deviations}), as it has been recently found in other
merging clusters (Markevitch \cite{mar10}).  
However, only an ICM
temperature analysis across the surface brightness discontinuity could
allow us to discriminate between the two possibilities, and with the
present dataset this analysis is not feasible because of the high soft
protons background.

A shock front generated during a cluster merger is expected to lie
perpendicular to the merging axis: see, e.g., the first known case of
1E 0657-56 (Markevitch et al. \cite{mar02}) and numerical simulations
(e.g. by Springel \& Farrar \cite{spr07}).  As for the clusters we
have already analyzed in the context of the DARC project, the shock
front in Abell 520 (Markevitch et al. \cite{mar05}) is perpendicular
to the NE-SW direction of the main merger (Markevitch et
al. \cite{mar05}, Girardi et al. \cite{gir08}); the likely shock front
detected in Abell 2219 (Million \& Allen \cite{mil09}) is
perpendicular to the SE-NW direction of the merger (Boschin et
al. \cite{bos04}); the shock front in Abell 2744 is perpendicular to
the SSE-NNW direction of the merger (Owers et al. \cite{owe11}),
somewhat a more southeasterly direction than that suggested by Boschin
et al. (\cite{bos06}).  As for A545, we note that the discontinuity is
roughly perpendicular to the bisecting line of the two likely merging
directions (i.e. towards NNW and towards NE, see Fig.~\ref{figk2z}).
This finding reminds the case of Abell 2345 -- hereafter A2345 -- 
where the western relic, likely generated by a shock, is found
to be perpendicular to the bisecting of the two merging directions
(Boschin et al. \cite{bos10}).  Note that, to date, no shock is
  detected in A2345 and we must rely on the position of the relic
  itself. Indeed, most relics are located far from the X--ray bright
  cluster regions, so it is rarely possible to check for the presence
  of a shock front in a X--ray image, and even more difficult to get a
  confirming temperature measurement, a likely exception being Abell
3667 by Finoguenov et al. (\cite{fin10}). Although A2345 and A545
  have both a complex merging structure which can help us to
  understand the position and geometry of the respective
  likely-shock/relic, the structure of these two clusters differ in
  two maior points. The subclusters are much closer in A545 than in
  A2345 (by roughly a factor two, cf. Fig.~\ref{figk2z} with Fig.~7 of
  Boschin et al. \cite{bos10}) suggesting a different merging age and,
  indeed, we below give a lower estimate for the time elapsed since the
  collision with respect to A2345.  Second, the main subcluster is
  richer -- with respect to other subclusters -- in A545 than in A2345
  (cf. the number of assigned galaxies by 2D DEDICA,$N_{\mathrm S}$
  in Table~2 of Boschin et al. \cite{bos10}).

Summarizing, we find that A545 is formed by three, likely four,
    subclusters and infer two directions for the accretion path
    (NNW-SSE and NW-SE directions).  We determine the (projected)
    position of the subclusters and give some indications about their
    relative importance and the angle of view of the merger. These
    results overcome the gross-scale result of an elongation in the
    ICM (Buote \& Tsai \cite{buo96}). Moreover, Buote \& Tsai gave no
    information about the merge age and, indeed, also assuming a
    post-collision phase, the elongation of the X-ray emission is a
    long-lived phenomenon (up to 5 Gyr after the core passage; see
    Roettiger et al. \cite{roe96}). More precise information can be
    derived from the northern discontinuity we observe in the X-ray
    surface brightness, if confirmed to be due to a shock. In fact, we
    can estimate the velocity of the shock front $v_s$ from ${\cal
      {M}}=v_s/c_s$ (e.g., Sarazin \cite{sar02}), where ${\cal {M}}$
    is the Mach number of the detected shock -- typically ${\cal{M}}
    \sim 2$ in clusters with radio halos (Markevitch et
    al. \cite{mar10}) -- and $c_s$ is the sound speed in the preshock
    gas -- $c_s\sim 1150$ \ks from the thermal velocity (i.e. from the
    observed $kT_X$ in the $\beta_{\rm spec}=1$ assumption).  We
    obtain $v_s\sim 2300$ \kss.  When using the distance of the X-ray
    surface brightness discontinuity from the cluster center,
    $r_{\mathrm shock}\sim 0.45$ \hh, we estimate that $t=r_{\mathrm
      shock}/v_s\sim 0.2$ Gyr is the time elapsed since the collision.

\subsection{A545 and predictions of halo/relic models}
\label{discnew}

As in the case of other clusters of DARC study, A545 supports the
scenario that extended, diffuse radio emissions are associated to
dynamically disturbed clusters, where major merging phenomena are
occurring.

More quantitatively, we can use our results to re-examine A545 with
respect to the observed scaling relations among $P_{\mathrm 1.4GHz}$,
the halo radio power, and $M_{\mathrm H}$, the total cluster mass
contained within the size of the radio halo, $R_{\mathrm H}$, and the
virial radius (Cassano et al. \cite{cas07}).  Rescaling the A545 mass
to $R_{\mathrm H}=380$ \kpc (Cassano et al. \cite{cas07}), we obtain
$M_{\mathrm H}(<R_{\mathrm H}=0.38 \hhh)\sim (0.2$--$0.4)$
\mquii. This is larger than the value of 0.12 \mqui reported by
Cassano et al., large part of the difference being due to the small
value assumed for the X-ray temperature ($kT_{\mathrm X}=5.5$ keV by
David et al. \cite{dav93}; Cassano et al. \cite{cas06}).  The result
is that the position of A545 moves from under to above the $M_{\mathrm
  H}$--$R_{\mathrm H}$ relation and from the left to the right of the
$P_{\mathrm 1.4GHz}$-$M_{\mathrm H}$ relation (Figs.~8 and ~9 by
Cassano et al. \cite{cas07}). This example points out the importance
of re-examining global cluster dynamical quantities to fix
observational scaling relations, in particular since these relations
are claimed to be predicted on the basis of the re-acceleration model
(Cassano et al. \cite{cas07}). When looking at the $R_{\mathrm
  H}$--$R_{\mathrm vir}$ relation obtained through numerical
simulations (Donnert et al. \cite{don10}), A545 can be used, as other
clusters listed by Cassano et al. (\cite{cas07}), to reject the
proposed hadronic model where the energy density of the cosmic ray
protons is taken as a constant fraction of the thermal energy density
population, $X_{CR}$=const, but not to check other two variations of
the basic model, i.e. using a radius dependent $X_{CR}$ and 
flattening the simulated magnetic field to be $B\propto \sqrt{\rho}$
with $\rho$ the mass density, where more massive clusters would be
useful (see Fig.~10 of Donnert et al. \cite{don10}).

We can also consider A545 with respect to the relation between
$\alpha_{\mathrm halo}$, the spectral index of radio halo, and
$r_{\mathrm relic}$, the distance of relic (or shock) from the cluster
center.  This relation is expected in the unified halo-relic hadronic
model by Keshet (\cite{kes10}) according to the idea that relic/shock
clustercentric distance can be used as a proxy for the time elapsed
since the merger.  Assuming that the northern discontinuity we observe
in the X-ray surface brightness is really due a shock, we obtain
$r_{\mathrm relic}=r_{\mathrm shock}\sim 450$ \kpcc. Considering that
$\alpha_{\mathrm halo}>1.4$ (Bacchi et al. \cite{bac03}), A545 well
lies in the $r_{\mathrm relic}$--$\alpha_{\mathrm halo}$ diagram
presented in the literature and only containing, to date, ten clusters
(Keshet \cite{kes10}, Fig.~28).

$\ $

To better investigate the structure of A545 we asked (and were
approved after the submission of this paper AO-10 - ID.067467 - PI
S. De Grandi) other time to XMM-Newton.  We are also planning new
spectroscopical observations since for the study of the dynamics of
complex merging clusters three times more galaxies need to solve the
cluster structure (e.g., in A520 Girardi et
al. \cite{gir08}). Finally, due to our indications for a merger mostly
occurring in the plane of sky, we point out to the advantage of having
deep photometry and a weak lensing analysis for better determine the
mass of the subclusters.

\subsection{Star pile and merging scenario}
\label{disc2}

The radial velocity of the star pile as measured by S07 is $47\,100\pm
60$ \kss. We show that the star pile is well at rest within the
cluster ($\left<\rm{v}\right>=47373\pm$125 \kss) and has the same
velocity of the CBCG ($47 071\pm$26 \kss).  However, the star pile is not
symmetric with respect to the CBCG (see our Fig.~\ref{figottico} and
Fig.~1 in S07, where the CBCG is indicated by ``SGH1''), rather lies
at SW with respect to the CBCG.  The star pile itself is someway
elongated in the NE-SW direction. As for the three faint galaxies
embedded in the star pile, the central one {(galaxy ``g2'' in
  Fig.~\ref{figottico})} has a velocity lower by 1300 \ks than the
star pile velocity as found by S07, the western object (the
M32-like dwarf elliptical cE, galaxy ``g1'') has a velocity
lower by 750 \kss, no redshift is available for the eastern object
  (galaxy ``g3'').

The red color of the star pile (S07) and the absence of a clear cool
core -- the X-ray analysis did not show any evidence of cool core from
both the surface brightness profile and the temperature analysis --
disfavor the possibility that the star pile is recently created by
cooling of the intracluster medium and subsequent star formation.  The
star pile in A545 might rather be the debris of a tidal stripping or
be created through a ``dry'' merger, i.e., a merger where the
progenitor galaxies contain little or no gas such that little star
formation is triggered by galaxy merger (e.g., Bell et
al. \cite{bel06}; Liu et al. \cite{liu09}).  The geometry and
the velocity of the star pile suggests that it is someway related with
the CBCG. Thus a possible source for the star pile is obviously the
CBCG itself or rather the compact dwarf cE, likely stripped by
interaction with the CBCG. In fact, compact elliptical galaxies are though
to form through tidal stripping (Nieto \& Prugniel \cite{nie87};
Bekki et al. \cite{bek01}).

In A545 the star pile is likely connected with the formation process
of a bright BCG at the cluster center at expenses of other galaxies as
suggested by Merritt (\cite{mer84}), and 
shown in numerical simulations (e.g., Dubinsky \cite{dub98}; 
  Murante et al.  \cite{mur07}) and in observations (e.g., van
Dokkum \cite{van05}).  In particular, the luminosity of the star pile
estimated by S07 corresponds to $L_{r^{\prime}}\sim 1\times 10^{11}$
\lsun (using $V-r^{\prime}$ colors for E-type galaxy by Fukugita et
al. \cite{fuk95}), i.e. roughly the double of the luminosity of the
CBCG. The co-addition of the CBCG, the star pile, and cE will lead to
a galaxy brighter than/as bright as the BCG1.

Moreover, in the case of A545, the global environment seems to be
critical for the formation of a BCG, too.  In fact, the CBCG seems to
point towards the NE group with respect to the star pile and the star
pile is elongated in the same direction.  The relative high velocity
of cE suggests instead that it, if really close to the CBCG, is presently
on a radial orbit with respect to the main central group.  This agrees
with the picture that BCG formation is caused by galaxies falling
along radial orbits aligned with the accreting filaments that feed
cluster growth (e.g., Dubinski \cite{dub98}).  Thus in A545 both the
intracluster light and the following, likely formation of a dominant
galaxy seem to be related to an extreme dynamical event such as a
cluster merger.  This might explain why the intracluster light in A545
is so asymmetric with respect to the CBCG when comparing the star pile
with the results of simulations (e.g., Dubinski \cite{dub98}). 

According to theoretical scenarios the hierarchical formation on both
galaxy and cluster scales are expected. In particular, BCGs are
  suggested to grow, at least in large part, via merging with 
  the BCGs of subclusters that have fallen into the cluster,
  although the rate of grow of BCGs seems slower than that of 
  parent clusters (Lin \& Mohr \cite{lin04} and refs. therein).  The
  memory of a connection between BCGs and their parent clusters is
  likely imprinted on the alignment of BCGs with their hosts (Sastry
  \cite{sas68}; Binggeli \cite{bin82}; Niederste-Ostholt et
  al. \cite{nie10}). Recent, detailed studies of clusters hosting BCGs
  (or central dominant ``cD'' galaxies) with multiple-nucleus provide
  important evidence for a scenario where the BCG/cD formation is
  likely due to the cluster merger. In particular, in a few, very
  interesting cases, the clump of cold, very dense gas centered on the
  BCG shows a plume aligned with (one of) the merging axis (Abell 3266
  by Henriksen \& Tittley \cite{hen02}; Abell 521 by Ferrari et
  al. \cite{fer06}). This gas is explained as due to stripped material
  related to an early phase of the BCG/cD formation, as suggested by
  the asymmetry in the gas distribution.

A545 is the first case where the connection between the cluster merger
and the BCG formation is strongly pointed out by an asymmetric
  stripped stellar matter aligned with the merging direction.  The
  asymmetry of the star pile indicates that its origin is very recent
  since N-body simulations find that streams of intracluster light
  found in the core of the clusters, which are characterized by short
  dynamical times, are quickly destroyed with times $\lesssim 1$ Gyr
  (Rudick et al. \cite{rud09}).  Present photometric data do not show
  evidence of a multiple-nucleus in CBGC as, e.g., in Abell 521
  (Ferrari et al. \cite{fer06}) and Abell 3266 (Henriksen \& Tittley
  \cite{hen02}) supporting the idea we are looking at a very
  preliminary phase of the BCG formation. Another case of a very
asymmetric intracluster light is that of CL 0958+4702 at $z=0.39$ but
there is no published study about the internal dynamics of this
  cluster (Rines et al. \cite{rin07}). In this context, A545
represents a textbook cluster where to study the simultaneous
formation of a galaxy system and its brightest galaxy.

After the submission of this paper for publication, Salinas et
  al. (\cite{sal11}) have presented new deep longslit observations of
  the star pile using VLT/FORS2 and Gemini/GMOS revealing a very
  irregular velocity field with parts of the star pile being
  associated with the g3 galaxy and other parts which have significant
  velocity offsets to the cluster systemic velocity. This chaotic
  velocity field supports our above scenario of a recent formation of
  the star pile and, indeed, also Salinas et al.  suggests that the
  origin of the star pile is due to tidal stripping of some galaxy.
  However, their interpretation stress the importance of the central
  cluster potential and of the g3 galaxy as the most probable
  ``central galaxy'' of the cluster, not assigning any special role to
  the CBCG (``C'' galaxy for Salinas et al.).  Our interpretation
  stress the role of the CBCG and, indeed, looking at their Fig.~6
  (upper panel, alternatively cf. their Table~4 and Fig.~5), the
  velocity of the CBCG/C seems equally close or slightly closer to the
  star pile velocity than the velocity of g3. Finally, we note that
  Fig.~6 of Salinas et al. also reinforce our idea that the cE/g1 is
  the remnant nucleus of a stripped galaxy since the star pile shows a
  positive velocity gradient from west -- the region of the low-velocity 
  cE/g1 -- to east towards the CBCG.
 
\begin{acknowledgements}

We are in debt with Federica Govoni for the VLA radio image and with
Giuseppe Murante for useful discussions.  We thank the referee for
her/his remarks which allowed us to make the paper clearer and more
complete.  M.G. acknowledges financial support from ASI-INAF
I/088/06/0 grant. This work has been also supported by the Programa 
Nacional de Astronom\'\i a y Astrof\'\i sica of the Spanish Ministry of Science 
and Innovation under grants AYA2010-21322-C03-02, AYA2007-67965-C03-01 
and AYA2010-21887-C04-04. This publication is based on observations made on
the island of La Palma with the Italian Telescopio Nazionale Galileo
(TNG) and the Isaac Newton Telescope (INT). The TNG is operated by the
Fundaci\'on Galileo Galilei -- INAF (Istituto Nazionale di
Astrofisica). The INT is operated by the Isaac Newton Group. Both
telescopes are located in the Spanish Observatorio of the Roque de Los
Muchachos of the Instituto de Astrofisica de Canarias.  X-ray data
are based on observations obtained with XMM-Newton, an ESA science
mission with instruments and contributions directly funded by ESA
Member States and NASA.  This research has made use of the NASA/IPAC
Extragalactic Database (NED), which is operated by the Jet Propulsion
Laboratory, California Institute of Technology, under contract with
the National Aeronautics and Space Administration.

\end{acknowledgements}

\end{document}